\documentclass[aps,prd,twocolumn,superscriptaddress,nofootinbib]{revtex4-2}

\usepackage{color}
\usepackage[dvipsnames]{xcolor}

\newcommand{\todooptional}[1]{} %

\usepackage{siunitx}
\DeclareSIUnit \h {\ensuremath{\mathit{h}}}
\DeclareSIUnit \Msun {M_\odot}
\DeclareSIUnit \parsec {pc}
\DeclareSIUnit \deg {deg}

\usepackage{graphicx}	%
\usepackage{amsmath}	%
\usepackage{amssymb}	%

\usepackage{multirow} %

\usepackage{colortbl}
\usepackage{physics}
\usepackage[normalem]{ulem}

\def\aj{AJ}%
\def\apj{ApJ}%
\def\apjl{ApJ}%
\def\apjs{ApJS}%
\def\ao{Appl.~Opt.}%
\def\aap{A\&A}%
\def\mnras{MNRAS}%
\def\prd{Phys.~Rev.~D}%
\def\prl{Phys.~Rev.~Lett.}%
\def\nat{Nature}%
\def\jcap{JCAP}%
\def\physrep{Phys.~Rep.}%
\usepackage[colorlinks=true,allcolors=blue]{hyperref}

\usepackage[capitalise]{cleveref}

\usepackage{xspace}

\newcommand{\camb}{\textsc{CAMB}\xspace}
\newcommand{\ccl}{\textsc{CCL}\xspace}

\newcommand{\namaster}{\textsc{NaMaster}\xspace}
\newcommand{\websky}{\textsc{Websky}\xspace}
\newcommand{\velocileptors}{\textsc{velocileptors}\xspace}

\begin{document}

\title{The Atacama Cosmology Telescope: DR6 Gravitational Lensing and SDSS BOSS cross-correlation measurement and constraints on gravity with the \texorpdfstring{E\textsubscript{G}}{EG} statistic}

\author{Lukas Wenzl}
\email{ljw232@cornell.edu}
\affiliation{Department of Astronomy, Cornell University, Ithaca, NY, 14853, USA}

\newcommand{\Geneve}{Universit\'e de Gen\`eve, D\'epartement de Physique Th\'eorique et CAP, 24 Quai Ansermet, CH-1211 Gen\`eve 4, Switzerland}

\newcommand{\DAMTP}{DAMTP, Centre for Mathematical Sciences, University of Cambridge, Wilberforce Road, Cambridge CB3 OWA, UK}
\newcommand{\KavliCambridge}{Kavli Institute for Cosmology Cambridge, Madingley Road, Cambridge CB3 0HA, UK}

\author{Rui An}
\affiliation{Department of Physics and Astronomy, University of Southern California, Los Angeles, CA 90089, USA}

\author{Nick Battaglia}
\affiliation{Department of Astronomy, Cornell University, Ithaca, NY, 14853, USA}

\author{Rachel Bean}
\affiliation{Department of Astronomy, Cornell University, Ithaca, NY, 14853, USA}

\author{Erminia Calabrese}
\affiliation{School of Physics and Astronomy, Cardiff University, The Parade, Cardiff, Wales CF24 3AA, UK}

\author{Shi-Fan Chen}
\affiliation{School of Natural Sciences, Institute of Advanced Study, Princeton NJ 08540, USA}

\author{Steve~K.~Choi}
\affiliation{Department of Physics and Astronomy, University of California, Riverside, CA 92521, USA}

\author{Omar Darwish}
\affiliation{\Geneve}

\author{Jo~Dunkley}
\affiliation{Joseph Henry Laboratories of Physics, Jadwin Hall, Princeton University, Princeton, NJ, USA 08544}
\affiliation{Department of Astrophysical Sciences, Peyton Hall, Princeton University, Princeton, NJ USA 08544}

\author{Gerrit~S.~Farren}
\affiliation{\DAMTP}
\affiliation{\KavliCambridge}

\author{Simone Ferraro}%
\affiliation{Lawrence Berkeley National Laboratory, One Cyclotron Road, Berkeley, CA 94720, USA}
\affiliation{Berkeley Center for Cosmological Physics, Department of Physics,
University of California, Berkeley, CA 94720, USA}

\author{Yilun Guan}
\affiliation{Dunlap Institute for Astronomy \& Astrophysics, University of Toronto, 50 St. George St., Toronto ON M5S 3H4, Canada}

\author{Ian Harrison}
\affiliation{School of Physics and Astronomy, Cardiff University, The Parade, Cardiff, Wales CF24 3AA, UK}

\author{Joshua Kim}
\affiliation{Department of Physics and Astronomy, University of Pennsylvania, Philadelphia, PA, 19104, USA}

\author{Thibaut Louis}
\affiliation{Université Paris-Saclay, CNRS/IN2P3, IJCLab, 91405 Orsay, France}

\author{Niall MacCrann}
\affiliation{\DAMTP}
\affiliation{\KavliCambridge}

\author{Mathew S.~Madhavacheril}
\affiliation{Department of Physics and Astronomy, University of Pennsylvania, Philadelphia, PA, 19104, USA}

\author{Gabriela A. Marques }
\affiliation{Fermi National Accelerator Laboratory, P. O. Box 500, Batavia, IL 60510, USA }
\affiliation{Kavli Institute for Cosmological Physics, University of Chicago, Chicago, IL 60637, USA}

\author{Yogesh Mehta}
\affiliation{School of Earth and Space Exploration, Arizona State University, 781 Terrace Mall, Tempe, AZ 85287, U.S.A.}

\author{Michael D. Niemack}
\affiliation{Department of Physics, Cornell University, Ithaca, NY 14853, USA}
\affiliation{Department of Astronomy, Cornell University, Ithaca, NY, 14853, USA}

\author{Frank J. Qu}
\affiliation{\DAMTP}
\affiliation{\KavliCambridge}

\author{Neelima Sehgal}
\affiliation{Physics and Astronomy Department, Stony Brook University, Stony Brook, NY 11794}

\author{Shabbir Shaikh}
\affiliation{School of Earth and Space Exploration, Arizona State University, 781 Terrace Mall, Tempe, AZ 85287, U.S.A.}

\author{Blake D.~Sherwin}
\affiliation{\DAMTP}
\affiliation{\KavliCambridge}

\author{Crist\'obal Sif\'on}
\affiliation{Instituto de F\'isica, Pontificia Universidad Cat\'olica de Valpara\'iso, Casilla 4059, Valpara\'iso, Chile}

\author{Alexander van Engelen}
\affiliation{School of Earth and Space Exploration, Arizona State University, 781 Terrace Mall, Tempe, AZ 85287, U.S.A.}

\author{Edward J. Wollack}
\affiliation{NASA Goddard Spaceflight Center, 8800 Greenbelt Rd, Greenbelt, MD 20771, USA}

\date{\today}

\begin{abstract}
We derive new constraints on the $E_G$ statistic as a test of gravity, combining the CMB lensing map estimated from Data Release 6 (DR6) of the Atacama Cosmology Telescope with SDSS BOSS CMASS and LOWZ galaxy data. We develop an analysis pipeline to measure the cross-correlation between CMB lensing maps and galaxy data, following a blinding policy and testing the approach through null and consistency checks. By testing the equivalence of the spatial and temporal gravitational potentials, the $E_G$ statistic can distinguish $\Lambda$CDM from alternative models of gravity. We find $E_G (z_{\rm eff} = 0.555) = 0.31^{+0.06}_{-0.05}$ for ACT and CMASS data at 68.28\% confidence level, and $E_G (z_{\rm eff} = 0.316) = 0.49^{+0.14}_{-0.11}$ for ACT and LOWZ. Systematic errors are estimated to be 3\% and 4\% respectively. Including CMB lensing information from \textsl{Planck} PR4 results in $E_G (z_{\rm eff} = 0.555) = 0.34^{+0.05}_{-0.05}$ with CMASS and $E_G (z_{\rm eff} = 0.316)= 0.43^{+0.11}_{-0.09}$ with LOWZ. These are consistent with predictions for the $\Lambda$CDM model that best fits the \textsl{Planck} CMB anisotropy and SDSS BOSS BAO, where $E_G^{\rm GR} (z_{\rm eff} = 0.555) = 0.401\pm 0.005$ for CMB lensing combined with CMASS and $E_G^{\rm GR} (z_{\rm eff} = 0.316) = 0.452\pm0.005$ combined with LOWZ. We also find $E_G$ to be scale independent, with $\textrm{PTE} >5\%$, as predicted by general relativity. The methods developed in this work are also applicable to improved future analyses with upcoming spectroscopic galaxy samples and CMB lensing measurements.

\end{abstract}

\maketitle

\section{Introduction}\label{sec:introduction}

While the current standard model of cosmology, $\Lambda$CDM, fits a wide range of observational constraints (e.g. \citep{Alam2017_DR12_cosmo_analysis,PlanckCollaboration2018,Freedman2019,Choi2020,Aiola2020,Madhavacheril2023}), the explanation behind the observed magnitude of the cosmological constant, $\Lambda$, remains unresolved \citep{Weinberg1989}. This has motivated the proposal of several alternative explanations, including a novel type of energy, called dark energy, and gravitational models including $f(R)$ gravity \citep{Carroll2004}, DGP \citep{Dvali2000}, Chameleon gravity \citep{Khoury2004}, TeVeS \citep{Bekenstein2004} and others (for a review see \citet{Clifton2012}). These modified gravitational models are generally designed to match the observed homogeneous accelerated expansion of the late-time universe and employ screening mechanisms so that they match general relativity (GR) on solar system scales \citep{Jain2010}. However, they do typically predict deviations from GR in the equations of motion affecting the growth of density fluctuations that seed large-scale structure (LSS) in the Universe \citep{Huterer2015}. Probing the growth of large-scale structure in the Universe, therefore, offers a powerful way to test gravity, allowing the construction and measurement of discriminating tests between $\Lambda$CDM and alternative models of gravity (see e.g. \citep{Jain2008,Jain2010,Ishak2019,Hou2023}). 

Photons from the Cosmic Microwave Background (CMB) get gravitationally lensed along the line of sight through the intervening gravitational potential formed from the large-scale structure of the Universe \citep{Lewis2000}. This CMB lensing effect can be detected in the CMB anisotropies through its characteristic effect of coupling different scales which would otherwise be independent from each other. CMB lensing maps have been created for recent CMB measurements from the \textsl{Planck} satellite \citep{Carron2022}, the Atacama Cosmology Telescope (ACT) \citep{Darwish2021,Qu2023} and the South Pole Telescope (SPT) \citep{Story2015,Pan2023}. The newest ACT CMB lensing map constructed based on the Data Release 6 (DR6) CMB dataset covers a total of $9,400~ \textrm{deg}^2$ and offers significantly higher resolution and signal-to-noise than previous maps based on \textsl{Planck} data \citep{Qu2023}.  

The cosmic volume probed by CMB lensing along the line of sight overlaps with the volumes probed by galaxy surveys. This allows measurements of the correlation between CMB lensing and galaxy clustering which serves as an opportunity to break degeneracies between the amplitude of fluctuations and the galaxy bias for galaxy clustering alone. The Sloan Digital Sky Survey III Baryon Oscillation Spectroscopic Survey
(SDSS BOSS) DR12 contains two large spectroscopic galaxy samples, CMASS and LOWZ which may be used to study this correlation, the release used is the latest and final version of the data \citep{Eisenstein2011,Reid2016,Alam2017_DR12_cosmo_analysis}. The correlation has been previously measured with the smaller ACT DR4 CMB lensing map in \citet{Darwish2021,Marques2023}. Combining the SDSS BOSS galaxy catalogs with the ACT DR6 CMB lensing map offers the opportunity to derive new competitive constraints on the correlation between CMB lensing and galaxy clustering, independent of \textsl{Planck}, and allows new constraints on gravity. To demonstrate the accuracy of this correlation we perform a detailed suite of null and consistency tests to show that biases from systematic effects like extragalactic foregrounds and magnification bias are mitigated and accurately accounted for. For this work, we focus on using these cross-correlation measurements for constraints on gravity and make them available for further analysis.

The measurement of this cross-correlation, combined with previous analyses of the galaxy data in \citet{Wenzl2024_EGestimator}, allow us to constrain the so-called $E_G$ statistic proposed in \citet{Zhang2007}. This statistic builds a discriminating test of gravity by comparing the divergence of the peculiar velocity field $\theta$ with a measurement of gravitational lensing $\nabla^2 (\psi -\phi)$. The proposed ratio compares the two sides of the generalized Poisson equation and tests the equivalence of the spatial and temporal gravitation potentials, directly testing the predicted equations of motion for GR \citep{Pullen2015,Leonard2015}. The estimator is designed to cancel the effect of galaxy bias on linear scales. To sample the velocity field a redshift space distortion (RSD) analysis of the galaxy sample is used to translate the measured galaxy clustering to velocities using the RSD parameter $\beta$. On linear scales and through matter conservation this parameter relates the divergence of the peculiar velocity field $\theta$ with the overdensities of galaxies $\delta_g$ as $\theta = \beta \delta_g$. The RSD analysis requires accurate redshifts and therefore this estimator is especially applicable to spectroscopic galaxy samples like those from SDSS BOSS. The first measurement of the $E_G$ statistic used an estimator based on galaxy weak lensing and galaxy clustering measurements \citep{Reyes2010}. In \citet{Pullen2015} an estimator for the $E_G$ statistic using CMB lensing and galaxy clustering was proposed. This was expanded upon and revised in \citet{Wenzl2024_EGestimator} to increase the overall accuracy and model several key systematic effects. We will use this revised estimator to measure $E_G$ in this work.

Previous analyses have investigated the $E_G$ statistic using CMB lensing information from \textsl{Planck} and the SDSS BOSS samples \citep{Pullen2016,Singh2019,Wenzl2024_EGestimator}. Some of these analyses claimed a significant tension with the $\Lambda$CDM prediction \citep{Pullen2016} while some found the measurement to be consistent with $\Lambda$CDM within statistical expectations \citep{Singh2019,Wenzl2024_EGestimator}, motivating further investigations. By leveraging independent CMB lensing measurements based on ACT observations we can derive new constraints providing a powerful consistency check. Additionally, the two lensing datasets are highly complementary: the \textsl{Planck} lensing information covers a larger sky area while the ACT dataset has a significantly higher angular resolution. By analyzing the two datasets using the same pipeline, and after checking for consistency between them, we derive a combined constraint resulting in the highest signal-to-noise constraint on $E_G$ with CMB lensing and galaxy data to date. We will put our results in context with constraints on $E_G$ from the literature at higher redshift using SDSS quasars \citep{Zhang2021} and ones using galaxy weak lensing instead of CMB lensing \citep{Reyes2010,Blake2016,delaTorre2017,Alam2017,Amon2018,Singh2019,Blake2020}. Careful investigations of the currently available datasets represent important preparatory work to develop and validate analysis pipelines applicable to upcoming datasets which will allow significantly improved constraints \citep{Pullen2015}. These include the Simons Observatory (SO) \citep{SimonsObservatory2019} and CMB-S4 \citep{Abazajian2016} for measurements of CMB lensing and for future spectroscopic samples such as from the Dark Energy Spectroscopic Instrument (DESI) \citep{DESICollaboration2016} and  \textsl{Euclid}.

In \cref{sec_background} we summarize the background for this work, giving definitions of the angular power spectra and the $E_G$ statistic and estimator used in this work. In \cref{sec_data} we describe the datasets used and in \cref{sec_blinding} the blinding policy followed during development. In \cref{sec_cross_corr_measrement} we describe how the analysis pipeline for the cross-correlation measurement between ACT DR6 CMB lensing and the SDSS BOSS galaxy samples is validated and applied. In \cref{sec_EG_measurement} we use the measurements to derive new constraints on the $E_G$ statistic and in \cref{sec_conclusion} we give a conclusion of the analysis.

\section{Background} \label{sec_background}

\subsection{Angular power spectra} %

Angular power spectra $C_\ell^{A B}$ are projections of a 3D power spectrum onto the sky between two observable tracers $A, B$ that are sensitive to specific distances described by kernels $W_A(\chi)$ and $W_B(\chi)$. Under the Limber approximation \citep{Limber1953} the angular power spectrum is given by
\begin{align}
    C_\ell^{A B} &\equiv \int \dd \chi \frac{W_A(\chi) W_B(\chi)}{\chi^2} P_{A B}\qty(k=\frac{\ell+1/2}{\chi}, z(\chi)),
\end{align}
where $\chi$ is the comoving distance and $P_{A B}(k,z)$ the 3D power spectrum the observations are sensitive to. 

Of interest for this work is the angular cross-correlation of CMB lensing with galaxy clustering and the angular auto-correlation of galaxy clustering. They are given by
\newcommand\GRequl{\mathrel{\stackrel{\makebox[0pt]{\mbox{\normalfont\tiny (GR)}}}{=}}}
\begin{align}
C^{\kappa g}_\ell &=\int \dd{z}  \frac{\hat W_{\kappa}(z) W_{g}(z)}{\chi^{2}(z)} P_{ \nabla^2 (\psi - \phi) g}\qty(k=\frac{\ell+1/2}{\chi(z)}, z)  \\ 
&\GRequl\int \dd{z}  \frac{W_{\kappa}(z) W_{g}(z)}{\chi^{2}(z)} P_{\delta g}\qty(k=\frac{\ell+1/2}{\chi(z)}, z), \\
C^{gg}_\ell &=\int \dd{z} \frac{H(z)}{c} \frac{W_{g}^2(z)}{\chi^{2}(z)} P_{gg}\qty(k=\frac{\ell+1/2}{\chi(z)}, z),
\label{eq:cl}
\end{align}
where $g$ refers to clustering and $\kappa$ refers to CMB lensing. Note here $W_g(\chi) = W_g (z) H(z) / c$.  Under the linear galaxy bias approximation used throughout this work, we have $P_{\delta g} = b P_{\delta \delta}$ and  $P_{gg} = b^2 P_{\delta \delta}$ where $P_{\delta \delta}$ is the matter power spectrum. Each tracer we consider has a specific kernel function $W(z)$ that characterizes the distances to which it is sensitive. The general lensing kernel for a source at redshift $z_{\rm S}$ is given by 
\begin{align}
\hat W_{\kappa} (z, z_{\rm S}) &= (1+z)\chi(z)\left(1- \frac{\chi(z) }{ \chi ( z_{\rm S})}\right), \\
W_{\kappa}\left(z, z_{\rm S}\right)&=\frac{3 H^2_{0} \Omega_{\rm m,0} }{2c^2}\hat W_{\kappa} (z, z_S).
\label{eq:lensing_kernel_fixed_source}
\end{align}
For CMB lensing the source is the surface of the last scattering $z_{*}$, so we have $\hat W_{\kappa}(z) \equiv \hat W_{\kappa}\left(z, z_{*}\right)$, $W_{\kappa}(z) \equiv W_{\kappa}\left(z, z_{*}\right)$. For the fiducial cosmology, we have $z_{*}=1089$. 

For galaxy clustering, the kernel can be modeled as
\begin{align}
    W_{g}\left(z \right)= \frac{\dd N}{\dd z},
    \label{eq:Wg}
\end{align}
where $\frac{\dd N}{\dd z}$ is the normalized galaxy redshift distribution accounting for all weights applied to the data. Here $N(z)$ refers to the cumulative weight of galaxies in the sample with redshifts up to $z$.

The observed galaxy clustering also has a contribution from a lensing signal of the foreground gravitational potential. This so-called magnification bias is an important systematic for the angular cross-power spectrum with lensing tracers and also affects the angular auto-power spectrum \citep{Krolewski2020,Maartens2021,vonWietersheimKramsta2021,Duncan2022,EuclidCollaboration2022_magnification_bias,ElvinPoole2023,Wenzl2023magbias}. The observed local number density of galaxies gets modulated through local weak lensing $\kappa$ as
\begin{equation}
    \frac{\Delta n}{n} = 2 (\alpha - 1) \kappa, \label{eq:general_definition_alpha}
\end{equation}
where $\alpha$ is a sample-specific parameter that can be estimated from the photometric selection of the galaxy samples \citep{Wenzl2023magbias}.

Including the magnification bias and assuming GR, the observed angular power spectra estimated from data are given as
\begin{align}
    \hat C^{\kappa g}_\ell& = C^{\kappa g}_\ell + \int \frac{d\chi}{\chi^2} W^\mu(\chi) W^\kappa (\chi) P_{\delta \delta}, \\ %
    \hat C^{gg}_\ell &= C^{gg}_\ell + 2 \int \frac{d\chi}{\chi^2} W^\mu(\chi) W^{g}(\chi) P_{\delta g} \nonumber \\ %
    &+ \int \frac{d\chi}{\chi^2} W^\mu(\chi) W^\mu(\chi) P_{\delta \delta}, %
\end{align} 
where the kernel for the magnification bias is given by 
\begin{align}
    W^\mu(\chi) = \int_{z(\chi)}^{\infty} dz'\,
      2 \,\left(\alpha(z') - 1 \right)\ \frac{dN(z')}{dz'}\ W^\kappa(z(\chi),z'). 
    \label{eq:zdep_wmu}
\end{align}

We calculate the theoretical predictions for $\hat C_\ell^{\kappa g}$ and $\hat C_\ell^{gg}$ for the fiducial cosmology in this work using \ccl\footnote{\url{https://github.com/LSSTDESC/CCL}} \citep{Chisari2019} and the underlying nonlinear 3D matter power spectrum using \camb \citep{Lewis2000,Howlett2012}. %

\subsection{Gravity Statistic \texorpdfstring{E\textsubscript{G}}{EG}} \label{sec_EG_gravity_statistic}

The $E_G$ gravity statistic tests the Poisson equation as predicted for GR by comparing the divergence of the peculiar velocity field, $\theta$, with gravitational lensing $\nabla^2 (\psi -\phi )$. Under the assumption of isotropy, homogeneity, and flatness, the universe can be described on large scales with the perturbed Friedmann-Lema\^itre-Robertson-Walker (FLRW) metric $\dd s^2 = (1+ 2\psi) \dd t^2 - a^2 (1+2\phi) \dd \textbf{x}^2$. Here $a(t)$ is the scale factor, with $a=1$ today, $\psi$ describes perturbations in the time component, and $\phi$ perturbations in the spatial component \citep{Bardeen1980}. The $E_G$ statistic is defined as \citep{Zhang2007}
\begin{align}
    E_G (k, z) \equiv \left[ \frac{\nabla^2 (\psi -\phi ) }{3 H_0^2 (1+z) \theta} \right]_{k}, \label{eq:EG}
\end{align}
where $H_0$ is the Hubble constant today and $k$ indicates scales in Fourier space.

In GR on linear scales, we have $\theta =  f\delta$, where $\delta$ is the total matter density contrast and $f=\dd \ln \delta/\dd\ln a$ is the logarithmic growth rate. In GR the statistic in \cref{eq:EG} is given by \citep{Zhang2007} %
\begin{align}
    E_G^{\rm GR} (z) = \frac{\Omega_{\rm m,0}}{f(z)},
\end{align}
where $\Omega_{\rm m,0}$ is the fractional matter density today compared to the critical density. In $\Lambda$CDM specifically, the growth rate is, to high accuracy, directly given by $f(z) = \Omega_{\rm m}(z)^{0.55}$ \citep{Linder2008_growth}, so that the $E_G$ statistic can be predicted from a measurement of $\Omega_{\rm m,0}$. In GR more broadly the growth rate is generally scale-independent on linear scales which represents another testable prediction of the theory. Alternative descriptions of gravity like $f(R)$ gravity \citep{Carroll2004,Hu2007} and Chameleon gravity \citep{Khoury2004}, while matching the background expansion of the universe, predict deviations in the growth history and therefore the $E_G$ statistic. In general, both the expected value of $E_G$ changes and scale dependence can be introduced in these alternative frameworks (see e.g. \citep{Zhang2007,Pullen2015}). 

The GR-based predictions can be compared to a measurement from data using a lensing tracer and a tracer of the non-relativistic velocities. For this work, we use a measurement of CMB lensing as the lensing tracer and we estimate the velocities from galaxy clustering and an RSD analysis. We use the harmonic space $\hat E_G$ estimator introduced in \citet{Wenzl2024_EGestimator} that builds on previous work by \citet{Pullen2015}. The estimator is given as\footnote{For simplicity, we drop the star in the notation of $\hat C_\ell^{\kappa g*}$ as given in \citep{Wenzl2024_EGestimator}. We apply the reweighting for the cross-correlation throughout to match the effective redshift of the measurements.}
\begin{align}
     \hat E_G^\ell (z_{\rm eff}) &\approx \Gamma_\ell (z_{\rm eff}) \frac{\hat C_\ell^{\kappa g}}{\beta \hat C_\ell^{gg}} \label{E_G_estimator},\\
    \Gamma_\ell (z_{\rm eff}) &\equiv C_\alpha^{\ell} \frac{2 c H(z_{\rm eff}) }{3 H_0^2} \int \dd z \frac{W_{g}^2(z)}{\hat W_\kappa (z)},
\end{align}
where $z_{\rm eff}$ is the effective redshift of the measurement, $H(z_{\rm eff})$ is the Hubble parameter at the effective redshift and $C_\alpha^{\ell}$ is a correction for the effect of magnification bias that is described in detail in \cref{sec:magnification_bias} and is based on the results of  \citep{Wenzl2023magbias,Wenzl2024_EGestimator}. The estimator combines the observed angular auto-power spectrum $\hat C_\ell^{gg}$ and $\beta$ from an RSD analysis of the galaxy sample with the angular cross-power spectrum with CMB lensing $\hat C_\ell^{\kappa g}$. 

The effective redshift of all three observables is matched by applying a reweighting scheme to the galaxies only when calculating the cross-correlation given by
\begin{align}
    w_\times (z) = \frac{\dd N}{\dd z} \frac{1}{\hat W_\kappa(z) I} \label{cross_reweighting},\\
    I = \int \dd z \frac{W_{g}^2 (z)}{\hat W_\kappa (z)}.
\end{align}
A new kernel for the galaxy sample to incorporate this additional reweighting is given by 
\begin{align}
W^*_g = \frac{\dd N^*}{\dd z} = \frac{\dd N}{\dd z} w_\times (z),
\label{eq:Wgstar}
\end{align}
and the effective redshift of the measurement by
\begin{align}
    z_{\rm eff} &= \frac{\int \dd z \, \chi^{-2} \hat W_\kappa(z) W^*_{g} (z) z}{\int \dd z \, \chi^{-2} \hat W_\kappa(z) W^*_{g} (z) }. \label{eq_effective_redshift}
\end{align}

In recent work, we demonstrated the accuracy of this estimator, before considering astrophysical systematics, at the level of
\begin{align}
\frac{E_G^{\rm GR} (z_{\rm eff})} {\hat E_G^\ell (z_{\rm eff})} - 1 < 0.3\%, 
\end{align}
for CMB lensing combined with the SDSS CMASS and LOWZ galaxy samples \citep{Wenzl2024_EGestimator}. Astrophysical effects like galaxy bias evolution with redshift and magnification bias could limit the overall accuracy. Galaxy bias evolution with redshift is difficult to correct for since it is degenerate with the cosmological signal of interest. In \citet{Wenzl2024_EGestimator} a $\Lambda$CDM fit was used to constrain the size of this systematic bias and it was concluded that for CMB lensing combined with CMASS and LOWZ, the systematic error budget is 1\% and 2\% respectively. For the present analysis, this is well within the statistical uncertainty and we account for it in our overall systematic error budget. For future surveys with larger constraining power, narrower bins in redshift can be used to reduce the impact of this systematic. Another systematic effect is magnification bias which can be estimated based on the photometric selection of the galaxy samples and corrected for in the analysis. The effect partially cancels for $E_G$ as the cross- and auto-correlations are affected in the same direction. The effect is cumulative along the line of sight and therefore more relevant for higher redshift samples: for LOWZ the magnification bias is negligible, but for CMASS a 2\textendash 3\% correction is needed which we discuss in detail in \cref{sec:magnification_bias}. The uncertainty of the correction is negligible compared to the other systematic error budgets and the statistical uncertainty \citep{Wenzl2024_EGestimator}.

\section{Data} \label{sec_data}

\subsection{Galaxy catalog from SDSS} \label{sec_SDSS_data}

For this work, we use the final Data Release (DR12) of the Baryon Oscillation Spectroscopic Survey (BOSS) which was part of Sloan Digital Sky Survey III \citep{Eisenstein2011}. For this program a total of 1,198,006 galaxy spectra were obtained over 10,252 square degrees of the sky culminating in two large-scale structure galaxy catalogs for cosmological analysis: CMASS and LOWZ \citep{Reid2016,Alam2017_DR12_cosmo_analysis}\footnote{\url{https://data.sdss.org/sas/dr12/boss/lss/}}. 

For the CMASS catalog, we select the galaxies with redshifts in the range $ 0.43 < z < 0.7 $ and for the LOWZ galaxy catalog in the range $0.15 < z < 0.43$. We construct overdensity maps using the fractional coverage masks presented in \citet{Wenzl2024_EGestimator}. We use only pixels with a coverage of at least 60\% for the analysis and apply SDSS weights to account for observational systematics. The weights for each galaxy are given by
\begin{align}
    w_{\rm auto} &= w_{\rm FKP} (w_{\rm NOZ} + w_{\rm CP} -1)\cdot  w_{\rm SEEING} \cdot w_{\rm STAR} \\
    w_{\rm cross} &= w_{\rm auto}  w_\times (z)
\end{align}
where the FKP weights are the optimal weights used for RSD analysis to minimize the uncertainty on $\beta$, and $w_\times$ are the additional weights specific to the cross-correlation [\cref{cross_reweighting}]. The other weights account in order for redshift failures, close pairs, seeing, and stellar contamination \citep{Reid2016}. The maps based on $w_{\rm auto}$ are used for the angular auto-power spectra and the galaxies have a normalized and weighted redshift distribution $\frac{\dd N}{\dd z}$ [\cref{eq:Wg}]. The maps based on $w_{\rm cross}$ are used for the angular cross-power spectra and have a weighted redshift distribution $\frac{\dd N^*}{\dd z}$ [\cref{eq:Wgstar}]. These two weighting schemes ensure that both the angular cross- and auto-power spectrum are sensitive to the same effective redshift given by \cref{eq_effective_redshift}. The effective redshift for the CMASS sample is $z_{\rm eff} = 0.555$ and for the 
LOWZ sample is $z_{\rm eff} = 0.316$.

 The overdensity maps are then constructed as
\begin{align}
    \delta_i = \frac{n_i} {f_i \bar n} -  1, \label{eq:overdensitymap}
\end{align}
where $i$ refers to each pixel, $f_i$ is the fractional coverage based on the mask, $n_i = \sum_{g \in i} w_{\rm cross/auto}$ is the weighted galaxy count and $\bar n = 1/N_{\rm pix} \sum_i n_i / f_i$. The sum is over the pixels in the mask and $N_{\rm pix}$ is the number of pixels in the mask. We report sky fractions for the galaxy map as $f_{\rm sky} = \sum_i f_i$ and sky fractions for the overlap by summing only the $f_i$ where we have CMB lensing information. 

To create simulated galaxy map realizations and to estimate the analytic covariance we need an estimate of the shot noise in the data given as \citep{Nicola2021,Marques2023,Wenzl2024_EGestimator}:
\begin{align} 
 \tilde N^{\rm shot}_{\rm cross/auto} &= \frac{f_{\rm sky}}{n_{\rm eff, cross/auto}} \label{eq:shotnoise}, \\
  n_{\rm eff, cross/auto} &= \frac{\left( \sum_{g} w_{\textrm{cross/auto}, g}\right) ^2}{4\pi f_{\rm sky} \sum_{g} w_{\textrm{cross/auto}, g}^2}, \label{eq:neff}
\end{align}
where $n_{\rm eff, cross/auto}$ is the effective number density of galaxies and the sums are over the weights for all galaxies.

\begin{figure}
\includegraphics[width=\columnwidth]{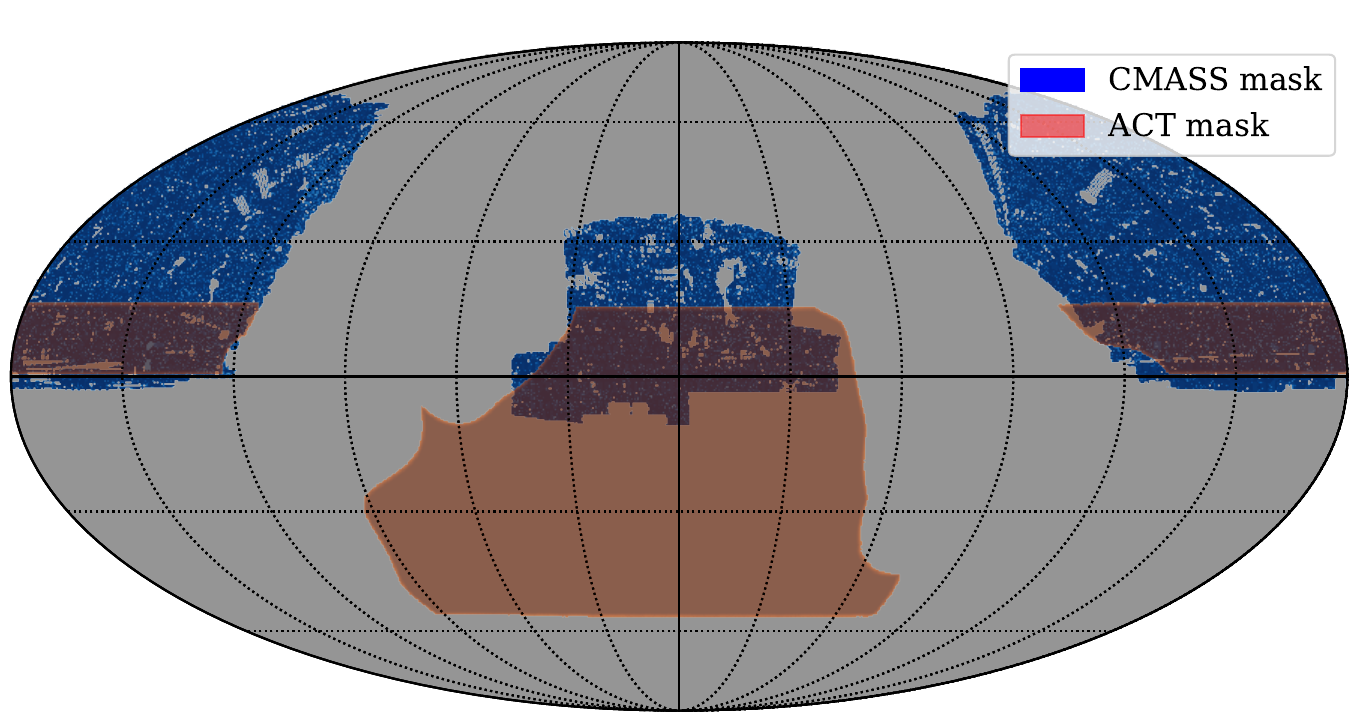}
\includegraphics[width=\columnwidth]{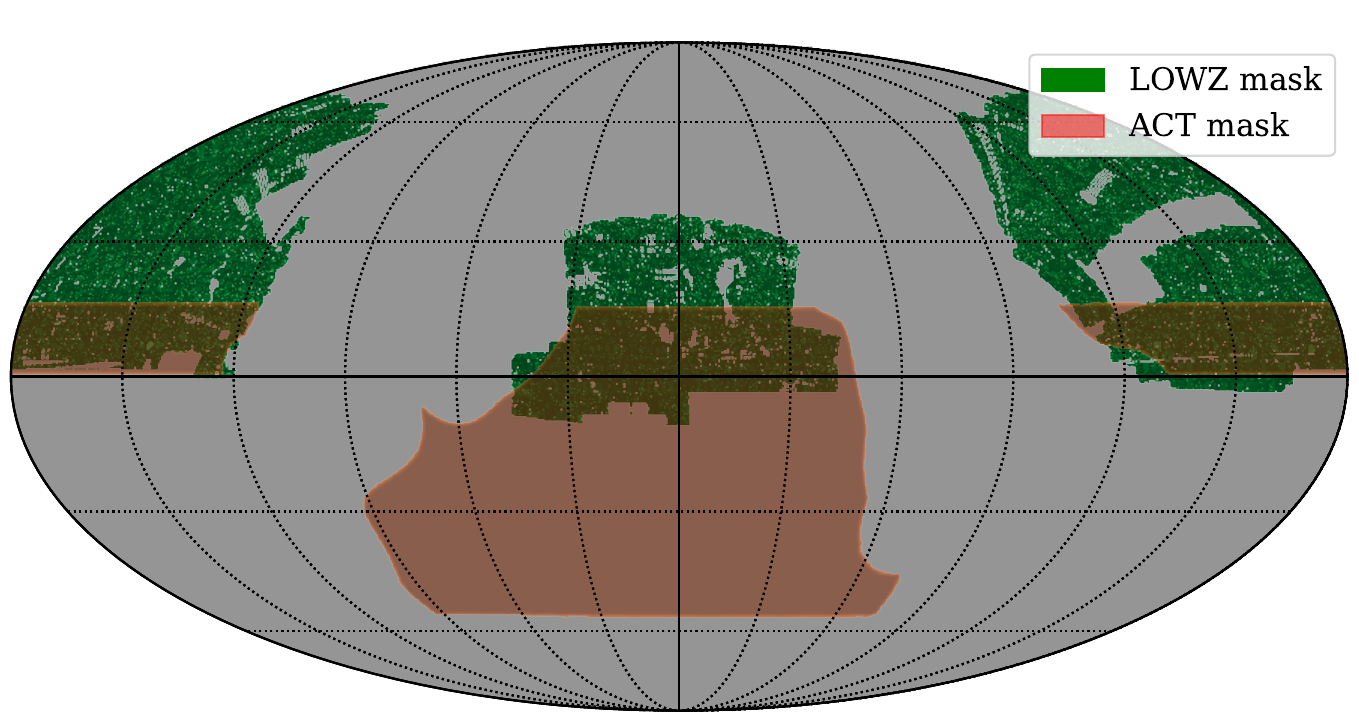}
\caption{Visualization of the overlap of the mask of the SDSS CMASS [upper, blue] and LOWZ [lower, green] samples with the mask of the ACT DR6 lensing map [orange] using Mollweide projections of the masks in equatorial coordinates (centered at dec and ra of 0 and west towards right). The hue of the colors indicates the coverage fraction for each pixel.}
\label{fig:masks}
\end{figure}

In \cref{fig:masks} we show the overlap of the mask of the ACT DR6 lensing map with the CMASS and LOWZ samples. The full CMASS map covers $f_{\rm sky} = 0.225$ and the region overlapping with the ACT DR6 lensing map has $f_{\rm sky} = 0.078$. For LOWZ the full map covers $f_{\rm sky} = 0.200$ and the overlap with ACT has $f_{\rm sky} = 0.074$.

\subsection{CMB lensing from the Atacama Cosmology Telescope} 
\label{sec_ACT_data}

The Atacama Cosmology Telescope (ACT) observed the CMB from 2007 to 2022 at high resolution. In 2016 the receiver was updated to the Advanced ACTPol \citep{Henderson2016}. The nighttime data from the f090 (around $90~\textrm{GHz}$)  and f150 (around $150~\textrm{GHz}$) bands from the years 2017-2021 measuring the CMB at a resolution of $0.5'$, form the basis for the DR6 release. 

The CMB lensing information based on temperature and polarization has already been extracted and published for the DR6 data \citep{Qu2023,Madhavacheril2023,MacCrann2023}.
The CMB lensing reconstruction is based on a bias hardened estimator \citep{Namikawa2013,Osborne2014,Sailer2020,Qu2023} which aims to reduce the effect of foreground contamination \citep{MacCrann2023}.

For this analysis, we use the ACT DR6 CMB lensing map and auxiliary data products presented in \citet{Madhavacheril2023}. In addition to the baseline lensing map, we use a range of data maps with alternative processing, the mask of the primary CMB observations, and the noise curve $N_\ell$ of the CMB lensing map. The alternative maps include CMB lensing reconstruction performed on the individual $90~ \textrm{GHz}$ and $150~ \textrm{GHz}$ maps as well as their difference using either temperature-only information or both temperature and polarization information. There is also a reconstruction that, instead of the baseline bias hardened estimator, uses CIB-deprojection as discussed in \citet{MacCrann2023}. This alternative map has a slightly more restrictive mask which we account for when making comparisons to the baseline maps.

In \citet{MacCrann2023} it was demonstrated, based on a range of tests including on the \websky CMB simulations \cite{Stein2019,Stein2020}, that the estimator used for the ACT DR6 CMB lensing map sufficiently mitigates the contamination of foregrounds in the analysis for the angular auto-power spectrum. From this alone, however, it does not necessarily follow that there is no contamination for angular cross-power spectra with LSS datasets as calculated in this work. We performed extensive testing for foreground contamination in the analysis (see \cref{sec_websky,sec_90150_nulltests,sec_CIB_deprojected}). Additionally, foreground contamination for the angular cross-power spectra with WISE galaxy samples was investigated in \citet{Farren2023} where no significant effect was found.

We apply a low pass filter defined by $\exp(-(\ell / \ell_{\rm max})^{20})$ using $l_{\rm max} = 1800$, to avoid signal bleed from small scales far outside the analysis range similar to comparable analyses \citep{White2022,Farren2023,Wenzl2024_EGestimator}. We also restrict the CMB lensing map to regions where the apodized CMB mask is above 0.6 to avoid dividing the maps by small values. 

\subsection{Simulated map realizations} \label{sec_simulated_map_realizations}

The ACT DR6 lensing data products include 400 simulated CMB lensing map realizations \citep{Madhavacheril2023}. The simulations consist of 400 simulated CMB realizations with a known CMB lensing realization applied \citep{Farren2023}. For each simulation, with the observational mask applied, the same reconstruction pipeline as used on the data was run which ensures that the simulations capture the full noise characteristics of the observed map. This makes them ideal for testing our analysis pipeline. The same has been done for the alternative maps used for the analysis. 

Using the input $\kappa$ realizations that were used to create the simulated ACT maps we create 400 simulated Gaussian galaxy overdensity map realizations for both CMASS and LOWZ that have the expected correlation with the simulated CMB lensing maps, the expected auto-correlation, and the same noise level as the observed galaxies given by the shot noise estimate. As done for the data we create two versions of the galaxy maps for the auto- and cross-correlation where for the cross-correlation we account for the shifted $\dd N^* / \dd z$ from the additional reweighting we do for the cross-correlation only [\cref{eq:Wgstar}]. This approach is described in detail in \citet{Wenzl2024_EGestimator}, building on previous work \citep{Kamionkowski1997,Farren2023}. We use linear galaxy biases of $b=2.14$ for CMASS and $b=2.04$ for LOWZ which were inferred by comparing the measured angular auto-power spectra of the galaxies to the amplitude of the input theory angular power spectra calculated for our fiducial cosmology. Before unblinding the data we used a preliminary value of $b=2$ \citep{White2011,Parejko2013}. 
We use the simulated maps to estimate the baseline covariance for the angular power spectra and show that our analysis accurately recovers the input angular-power spectra.

\subsection{CMB lensing measurement from Planck} \label{sec_Planck}

We also report combined constraints with \textsl{Planck} CMB lensing measurements. For this, we use the latest \textsl{Planck} PR4 data, also called NPIPE \citep{PlanckCollaboration_PR4_NPIPE}. We use the CMB lensing map based on the PR4 analysis presented in \citet{Carron2022}\footnote{\url{https://github.com/carronj/planck_PR4_ lensing}}. 

We apply the same data processing as described in \citet{Wenzl2024_EGestimator}: a low pass filter $\exp(-(\ell / \ell_{\rm max})^{20})$ with $l_{\rm max} = 1800$ in order to avoid signal bleed from small scales far outside the analysis range and an apodization to avoid sharp edges as described in \citet{White2022}. To allow combination with our other maps we rotate the \textsl{Planck} CMB lensing map to equatorial coordinates in harmonic space. 

To evaluate the covariance of the measurement we use a set of 480 simulated realizations of the CMB lensing maps \citep{PlanckCollaboration2020_sims,Carron2022}. %

\section{Blinding procedure} \label{sec_blinding}

The results presented in this work were derived as part of the ACT collaboration and followed the ACT DR6 lensing blinding guidelines.

During development, we blinded the ACT DR6 CMB lensing map with an unknown $\approx 10\%$ blinding amplitude (see \citet{Qu2023} for additional details on the blinding approach). We developed the null and consistency tests on the blinded map. We did not compare the measured curves to the theory during development. Furthermore, we finalized the $E_G$ estimator used in this work and the analysis choices like the binning scheme and cosmological ranges before unblinding. 

After confirming the accuracy of the pipeline we performed a step-wise unblinding procedure. First, we used the amplitudes of the $C_\ell^{gg}$ measurements to refine the galaxy bias values from their initial guesses of $b=2$ used during development. Then we recreated the galaxy simulations with the updated theory curves, reran the null and consistency tests on the unblinded maps, and confirmed they still show statistical consistency. The values for the null and consistency tests we show in the paper are the ones after unblinding. We use an estimator developed on SDSS and BOSS data from \citet{Wenzl2024_EGestimator}.

Finally, we unblined the ACT-based $E_G$ results and compared them to the GR predictions. After revealing the results we investigated the sensitivity of the results to various analysis choices and combined the two angular cross-power spectra to derive a combined $E_G$ constraint.

\section{Angular cross-power spectrum measurement} \label{sec_cross_corr_measrement}

\subsection{Analysis approach}

\subsubsection{Estimation of angular cross-power spectra} \label{sec:signal_estimation}

We measure the angular cross-power spectra $C_\ell^{\kappa g}$ from the map as the cross-variance between the two maps at a set of scales. The mask geometry couples different scales, which can be accounted for analytically. For this, we use the unified pseudo-$C_\ell$ framework provided by \namaster\footnote{\url{https://github.com/LSSTDESC/NaMaster}} \citep{Alonso2019}.
Additionally, with this approach, we combine multiple multipoles into bandpowers which are more computationally efficient and are less correlated. In this section, we label bandpowers by $p$ for readability but thereafter will return to label them as $\ell$, with $\ell$ being the effective scale that the given bandpower samples. The angular cross-power spectrum $C_p$ for maps $a$ and $b$ with partial sky coverage can be calculated from their masked spherical harmonics coefficients $a_{lm}$ and $b_{lm}$ as
\begin{align}
    C_p &= \sum_{p'} \left[ \mathcal{M}^{-1} \right]_{p p'} \tilde C_{p'} \label{eq:Cell_estimate} \\
    \tilde C_p &= \sum_{\ell \in p} \tilde \omega_\ell \frac{1}{2\ell +1} \sum_{m=-\ell}^{\ell} a_{\ell m} b_{\ell m}, \label{eq:Cell_estimate_justanafast}
\end{align}
where $\mathcal{M}$ is the binned coupling matrix, calculated based on the mask of both tracers. $\tilde \omega_\ell$  is the weight for each scale, in our case the inverse of the number of $\ell$ in the corresponding bandpower. To first order, the mode coupling matrix in \cref{eq:Cell_estimate} is rescaling the pseudo-$C_\ell$ by $f_{\rm sky}$, the covered sky fraction of the overlap between the two maps.

We account for the partial completeness of each pixel in our maps by rescaling them and therefore use the default \texttt{masked\_on\_input = False} when creating \namaster field objects. For the galaxy overdensity maps this rescaling is the inverse of the galaxy mask as described in \cref{eq:overdensitymap}.
However, for $\kappa$ maps, we use the inverse of the CMB mask squared which is an approximation of the mask of the lensing map. We test that this is a reasonable approximation for our purposes through our end-to-end pipeline test in \cref{sec_input_recovery}. 

Since the galaxy maps are constructed at finite resolution in real space without taking the sub-pixel coordinates into account they are affected by a pixel window function \citep{Marques2023}. We use map resolutions of nside=1024 throughout and account for the effect when calculating $C_\ell^{\kappa g}$ by including the pixel window function as an effective beam function in the galaxy field object.

We use a log-spaced binning scheme. For computational accuracy, since different scales are correlated, the computations are performed over a sufficiently wide range limited by our map resolution: $2 \leq \ell < 3071 = 3\,  \textrm{nside}-1 $. We limit the cosmological analysis to $\ell \geq 48$ throughout as ACT is not sensitive to the largest scales. The cutoff value is chosen to be consistent with previous analyses \citep{Wenzl2024_EGestimator,Chen2022,Pullen2016}. Additionally, we limit the cosmological range to scales $k< \SI{0.2}{Mpc^{-1}}$ where the linear assumptions underlying the $E_G$ statistic are valid. This represents a comoving real space cutoff at around $\SI{30}{Mpc}$. For the effective redshift of the galaxy samples this results in an upper cutoff of $\ell_{\rm max} = 420$ for CMASS and $\ell_{\rm max} = 233$ for LOWZ where the exact values are chosen to be edges in our binning scheme. The data allows constraining of the angular cross-power spectrum to smaller scales beyond the linear regime. We perform our consistency tests in an extended validation range  $40 \leq \ell < 922$. The exact edges of the log-spaced binning scheme over the full extended range are given by $\ell_{\rm edge} \in$ [40, 48, 59, 71, 87, 106,  129,  157,  191,  233, 283, 345, 420, 511,  622,  757,  922] where the bins are inclusive of the lower edge and exclusive of the upper edge. 

Since this approach to get an unbiased estimate of the power spectrum assumes the underlying power spectrum to be a step function matching the bandpowers, we also need to correct the theory $C_\ell$ for the bandpower binning. This is done by applying the coupling matrix, then binning and then applying the inverse of the binned coupling matrix \citep{Alonso2019}. We apply this throughout to compare theory spectra to data and for calculating the magnification bias correction in \cref{sec:magnification_bias}.

\subsubsection{Covariance estimation and consistency} \label{sec_covariance_estimation}

In this analysis, we use three methods for estimating the covariance of the measurement and confirm they are consistent with each other within the expected variation. Our baseline approach is to estimate the covariance based on simulations. We can compare this to an analytic estimate that does not suffer from noise and a data-based estimate using a jackknifing approach that is sensitive to any additional noise in the data not included in the modeling.

From our simulated maps, we can estimate the covariance by calculating the angular power spectra for each realization and estimating the variance between the simulations. The general form is given by
\begin{align} 
 \widehat {\textrm{Cov}}(X_\ell, Y_{\ell'}) = \frac{1}{N_{\rm sims}-1} \sum_{j=1}^{N_{\rm sims}} \left( X_\ell^{(j) }- \bar X_\ell\right) \left( Y_{\ell'}^{(j) }- \bar Y_{\ell'}\right),  \label{eq_cov_sim_based}
\end{align}
where for the covariance of the angular cross-power spectrum $X_\ell = Y_\ell = \hat C_\ell^{\kappa g}$ and $\bar X_\ell, \bar Y_\ell$ refer to the mean over all simulations. $N_{\rm sims}$ is the number of simulations, which is 400 in this case and the sum is over the angular power spectra calculated for each of the 400 realizations indexed by $(j)$. The same approach can also be used to estimate the covariance between $X_\ell = C_\ell^{\kappa g}$ and the angular auto-power spectrum $Y_\ell = C_\ell^{gg}$ as well as the covariance of $C_\ell^{gg}$ itself. For likelihood analyses, one generally needs the inverse of the covariance. To get an unbiased estimate for the inverse of a noisy covariance we apply \citet{Hartlap2007} corrections when inverting given by
\begin{align}
    \widehat {\textrm{Cov}}^{-1} \rightarrow \widehat {\textrm{Cov}}^{-1} \left(1 -\frac{N_d +1 }{N_{\rm sims}}\right) \label{eq:Hartlap},
\end{align}
where $N_d$ is the dimension of the covariance. We limit the covariance to the bins in the cosmological analysis range when inverting and apply Hartlap corrections throughout for simulation and jackknife-based covariances. 

In addition to the covariance estimated from simulated realizations, we also estimate the covariance analytically. For this we use the angular power spectra $C_\ell^{gg}, C_\ell^{\kappa g} $ and $C_\ell^{\kappa \kappa}$ for our fiducial cosmology, the masks of the CMB lensing and galaxy maps, an estimate of the shot noise in the galaxy map and an estimate of the noise $N_\ell$ in the CMB lensing map. %
For the mask of the CMB lensing map, we use the square of the mask of the primary CMB observations which is an approximation we confirm to be sufficiently accurate through our input recovery test (\cref{sec_input_recovery}). We use the method from \citet{GarciaGarcia2019} implemented in \namaster as the \texttt{NmtCovarianceWorkspace} class to estimate the Gaussian covariances. This gives us analytic estimates of the covariance of $C_\ell^{\kappa g}$ as well as the covariance for $C_\ell^{gg}$ and the cross-covariance between them. 

\begin{figure*}
\includegraphics[width=0.975\columnwidth]{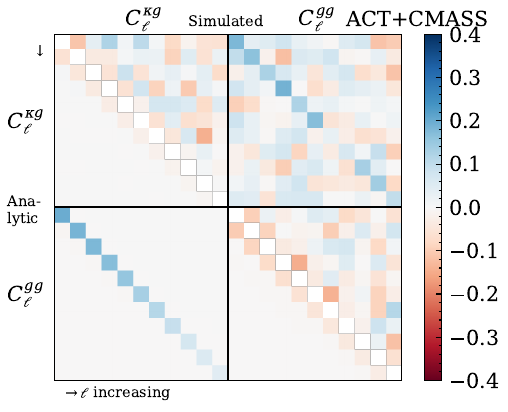}
\includegraphics[width=\columnwidth]{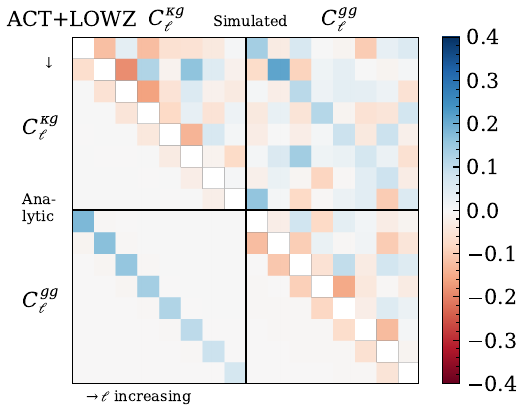}
\caption{Correlation matrices for the covariance matrices of the angular power spectra with the correlation shown in color from -0.4 (red) to 0.4 (blue) in $\ell$ bins from $\ell$= 40 to 922, as specified in \ref{sec:signal_estimation}. The diagonal of $1$ is removed for clarity. Shown is the full covariance for $C_\ell^{\kappa g}$ and $C_\ell^{gg}$ for [Left] ACT + CMASS and [Right] ACT + LOWZ for the bins in the cosmological analysis range. The upper triangles show the baseline simulation-based covariance and the lower triangle the analytic covariance.} 
\label{fig_covariances}
\end{figure*}

Finally, we also estimate the covariance of the measurements from the data directly using a jackknife approach to test if there are any additional effects in the data that were not included in our modeling. For jackknife estimation of the covariance for $C_\ell^{\kappa g}$ we split the overlapping area between the CMB lensing and galaxy maps into $N_{\rm jack}$ equal area patches. 
The covariance of the measurement can then be estimated by summing the variation in the estimate across all patches. The general form for the covariance between two angular power spectra $X_\ell, Y_\ell$ is given by
\begin{align}
    \widehat {\textrm{Cov}}(X_\ell, Y_{\ell'}) = \frac{N_{\rm jack}-1}{N_{\rm jack}} \sum_{j=1}^{N_{\rm jack}} \left( X_\ell^{(j) }- \bar X_\ell\right) \left( Y_{\ell'}^{(j) }- \bar Y_{\ell'}\right), \label{eq_cov_jack_based}
\end{align}
where for the angular cross-power spectra $X_\ell = Y_\ell = C_\ell^{\kappa g}$ and $C_\ell^{\kappa g (j)}$ is the angular cross-power spectra when each of the $j=1, N_{\rm jack}$ patches is individually removed from the maps. 
To capture the variance in the largest scales one generally chooses patch sizes so that the largest scales in the analysis are still captured \citep{Pullen2016}. Due to this we only use 16 patches limited by the size of the overlap between the ACT CMB lensing map and the SDSS galaxy samples. Because of this, the statistical uncertainty of this estimate is large compared to the other methods described above.

Furthermore using a jackknife approach systematically overestimates the variance in angular power spectra \citep{Marques2020,Mohammad2022}. Because of these limitations, we expect the jackknife errors to be on average a systematic overestimate of 20-30\% with large statistical variation. We check for our analysis that this is true on average and there are no outliers of the marginalized jackknife error being more than 50\% larger than our simulated and analytic estimates. Due to the limited amount of samples, we do not estimate the full covariance for the jackknife case, but we report the binned $E_G^{\ell}$ results for comparison to the other estimates. 

We compare the simulated covariance and analytic covariances showing the off-diagonals of the covariance as a correlation matrix. The correlation matrices are calculated from the covariances as $\textrm{Cov} (C_\ell^{XY}, C_\ell^{AB}) / (\sigma_\ell^{XY} \sigma_\ell^{AB}) $.  
The off-diagonals of the covariance are consistent between the two methods within the statistical uncertainty of the simulation-based estimate. The most relevant correlations are between auto- and cross-power spectra at the same scale and for neighboring bins. 

The amplitudes of the diagonals predicted by three different covariances approaches are also consistent as discussed in \cref{sec_cross_corr_measrement}.
As we will show in the comparison of the analytic and jackknife-based error estimates relative to the simulation-based estimates in the middle panels of \cref{fig:cross_corr_results}, we find that both estimates are within expectations. This gives us confidence to use our simulation and analytic approaches as the covariance estimates.

\subsubsection{Normalization correction} \label{sec_norm_correction}

As discussed in detail in \citet{Qu2023} the $k$-space filtering applied to the ACT DR6 maps causes a loss in power for the $\kappa$ map reconstruction. We apply a simulation-based Monte Carlo (MC) correction to our lensing map $\hat \kappa_{\ell m}$ to account for this loss in signal. 

We measure this MC-correction based on the set of input $\kappa$ maps sampled from our theory $C_\ell^{\kappa \kappa}$ compared with the $\hat \kappa$ reconstruction maps of CMB realizations with the corresponding input lensing maps applied, i.e. our set of 400 simulations. These reconstructions use the same pipeline as our measured CMB lensing map and are therefore affected in the same way. The MC-correction is defined as the inverse of a transfer function so it can be directly multiplied on our data maps. The functional form is
\begin{align} 
\textrm{MC-corr}(\ell) &\equiv \frac{\bar C_{\ell}^{\kappa_{g\rm -mask}, \kappa_{\rm CMB-mask^2}}  }{\bar C_{\ell}^{\kappa_{g\rm -mask}, \hat \kappa} } , \\
 \hat a_{\ell m}^{\kappa} &\rightarrow \textrm{MC-corr}(\ell )\cdot \hat a_{\ell m}^{\kappa}, \label{eq:MC_corr}
\end{align}
where $\kappa_{g\rm -mask}$ and $\kappa_{\rm CMB-mask^2}$ are the simulated input realizations of the lensing map, where the first one has the specific mask of the galaxy sample that we want to correlate with applied and the second one has the CMB mask squared applied. The restriction of the angular power spectrum to the overlap with the galaxy map considered ensures that we get the correction for the specific area relevant to the analysis. Using the square of the CMB mask an approximation of the mask for the derived lensing reconstruction. We confirm that this approximation is accurate for our purposes with our end-to-end pipeline test showing that we accurately recover the input spectra to the precision needed (\cref{sec_input_recovery}). Note it is numerically advantageous to average over our set of sims before taking the ratio. Furthermore, the correlations are calculated only using \cref{eq:Cell_estimate_justanafast}, since we do not want to account for the coupling matrix in our normalization correction to avoid correcting for it twice. We implement this calculation using \texttt{healpy.sphtfunc.anafast}.

For the full ACT map, the MC-correction is approximately 11\% in our cosmological range, increasing the measured $C_\ell$. This baseline correction is already applied to the published ACT DR6 lensing maps. We replace the correction with ones calculated for the specific area overall with CMASS and LOWZ to check for area dependence of the correction. For our specific areas considered, we find these corrections differ from the one for the full map by less than 1\%. 
For numerical stability, we extrapolate the MC-correction far outside our analysis range as a constant below $\ell = 20$ and above $\ell = 1000$ and in between smooth over 40 $\ell$.

This approach to the correction is agnostic to the measured amplitude of the CMB lensing signal, it only uses simulations and was finalized while the data was still blinded. Nonetheless, there are shortcomings as it, for example, assumes the shape of the MC-correction for the $\kappa$ auto-correlation is similar to that of the cross-correlation. While it is out of scope for this project, future work could further investigate the optimal handling of the transfer function. Further details about the normalization correction for the ACT DR6 lensing maps can be found in \citet{Qu2023,Farren2023}.

We apply the normalization corrections to all simulated and observed $\kappa$ reconstruction maps for our results and null tests. When comparing the North and South patches of SDSS we calculate the normalization correction separately for each. Since the normalization correction affects the amplitude of the cross-correlation and therefore $E_G$, we evaluate to what degree our results and conclusion depend on the accuracy of the normalization correction in \cref{sec_analysis_choices}.

\subsubsection{Magnification bias}\label{sec:magnification_bias}

Based on the results in \citet{Wenzl2023magbias} we use magnification biases of $\alpha_{\rm LOWZ} = 2.47 \pm 0.11$ for LOWZ and for CMASS we use 
\begin{equation}
    \alpha_{\rm CMASS}(z) = 2.71 + 8.78 (z - 0.55).
    \label{eq:cmass_zdep}
\end{equation}
For the latter, we assume a statistical uncertainty of 0.08 for the constant and 1.26 for the slope. Here we combine the statistical and systematic uncertainties conservatively by adding them.

We include the magnification bias terms in the theory curves for $\hat C_\ell^{\kappa g}$ and $\hat C_\ell^{gg}$ that enter the analysis as part of creating simulated map realizations, the covariance estimation and when plotting the theory curves together with the data. 

We also use the measured $\hat C_\ell^{\kappa g}$ including the magnification bias term as input for our $E_G$ estimate. The $E_G$ estimator (see \cref{E_G_estimator}) accounts for the magnification bias effect on the angular auto- and cross-power spectra via a correction term given as
\begin{align}
    C^\ell_\alpha &\equiv \frac{C_\ell^{\kappa g}}{\hat C_\ell^{\kappa g}} \frac{\hat C_\ell^{gg}}{C_\ell^{gg}},
\end{align}
where $\hat C_\ell^{\kappa g}$ and $\hat C_\ell^{gg}$ are the theory angular power spectra for our fiducial cosmology with the magnification bias contribution and $ C_\ell^{\kappa g}$ and $ C_\ell^{gg}$ without. 

This results in a negligible correction for $E_G$ with CMB lensing and LOWZ and a 2\textendash 3\% correction downwards in amplitude for the measurement of $E_G$ with CMB lensing and CMASS (See Fig. 2 of \citet{Wenzl2024_EGestimator} for a plot of the $C^\ell_\alpha $ values).

\subsection{Consistency and null tests} \label{sec_consistency_and_null_tests}

We perform a range of null and consistency tests on our angular cross-power spectrum analysis pipeline to ensure its accuracy. We validate the pipeline by showing that we accurately recover the input based on simulations (\cref{sec_input_recovery}). We further confirm our covariance estimates by considering the correlation of misaligned maps (\cref{sec_misaligned_corr}). We perform a range of tests that are sensitive to foreground contamination to the result to build confidence in the cross-correlation analysis (\cref{sec_websky,sec_90150_nulltests,sec_CIB_deprojected}). We show that the analysis is consistent between the SDSS North and South patches to justify combining them for the presented work (\cref{sec_north_v_south}). %

In this analysis, we assume angular power spectra are Gaussian distributed so that their uncertainty can be described with a covariance matrix. To test our analysis approach we perform a large set of statistical tests.

When testing an angular power spectrum measurement $C_\ell$ with covariance $\textrm{Cov}(C_\ell,{C_{\ell'}})$ against a null hypothesis we calculate $\chi^2$ values as
\begin{align}
    \chi^2 = \sum_{\ell \ell'} C_\ell \textrm{Cov}^{-1}(C_\ell,{C_{\ell'}}) C_{\ell'} .  
\end{align}
When testing two angular power spectra $C_\ell^X$, $C_\ell^Y$ for the hypothesis of consistency we calculate $\chi^2$ values as
\begin{align}
    \chi^2 = &\sum_{\ell \ell'} D_\ell \textrm{Cov}^{-1}(D_\ell,D_{\ell'}) D_{\ell'}, \\
    D_\ell = &C_\ell^X - C_\ell^Y ,\label{chi_sq_diff}\\
    \textrm{Cov}(D_\ell,D_{\ell'}) = &\textrm{Cov}(C_\ell^X,C_{\ell'}^X) +  \textrm{Cov}(C_\ell^Y,C_{\ell'}^Y) \nonumber \\ &-\textrm{Cov}(C_\ell^X,C_{\ell'}^Y) - \textrm{Cov}(C_\ell^Y,C_{\ell'}^X),
\end{align}
where $D_\ell $ is the difference between the observed angular power spectra and $\textrm{Cov}(D_\ell,D_{\ell'})$ the covariance for the difference. In some cases, the two measurements are independent, and therefore the cross-covariance can be neglected. However, in cases where the angular power spectra being compared are significantly correlated, it is crucial to account for the cross-covariance terms to not underestimate the $\chi^2$ value. Given the degrees of freedom (DOF) for the test, we calculate the probability of getting a $\chi^2$ value at least this large, referred to as probability to exceed (PTE). These PTE values are calculated as the integral of a $\chi^2$ distribution for the specific DOF from the measured $\chi^2$ value to infinity. Note that for Gaussian uncertainty a PTE value of 0.68 corresponds approximately to a $1\sigma$ difference and a PTE value of 0.05 approximately to $2\sigma$. For previous applications of PTE values see for example \citet{Qu2023,Farren2023,Wenzl2024_EGestimator}.

In the following paragraphs, we discuss the motivation and findings of individual null and consistency tests. The resulting PTE values are summarized in \cref{table_null_tests}. In \cref{sec_summary_of_PTE} we discuss the distribution of all PTE tests presented as part of this work.

\begin{table*}%
\resizebox{\textwidth}{!}{%
\begin{tabular}{|l|l|c|c|c|c|c|c|c|c|}
\cline{3-10}
\multicolumn{1}{c}{}&  & \multicolumn{4}{c|}{ACT $\times$ CMASS}  & \multicolumn{4}{c|}{ACT $\times$ LOWZ} \\ \hline
 \multirow{2}{*}{\begin{tabular}[c]{@{}l@{}}Consistency and\\ Null Tests\end{tabular}} & \multirow{2}{*}{Primary purpose} & \multicolumn{2}{c|}{\begin{tabular}[c]{@{}c@{}}Cosmo range \\ ($48 \leq \ell  < 420$)\end{tabular}} & \multicolumn{2}{c|}{\begin{tabular}[c]{@{}c@{}}Ext. valid. range \\ ($40 \leq \ell < 922$)\end{tabular}} & \multicolumn{2}{c|}{\begin{tabular}[c]{@{}c@{}}Cosmo range \\ ($48 \leq \ell  < 233$)\end{tabular}} & \multicolumn{2}{c|}{\begin{tabular}[c]{@{}c@{}}Ext. valid. range\\ ($40 \leq \ell < 922$)\end{tabular}} \\ \cline{3-10}
 &  & $\chi^2 / \textrm{DOF}$ & PTE & $\chi^2 / \textrm{DOF}$ & PTE & $\chi^2 / \textrm{DOF}$ & PTE & $\chi^2 / \textrm{DOF}$ & PTE \\ \hline \hline
Signal recovery on Sims & \begin{tabular}[c]{@{}l@{}}Pipeline, \\ $\textrm{mask}^2$ approx\end{tabular} & 6.2/11 & 0.86 & 10.7/16 & 0.83 & 9.2/8 & 0.32 & 24.0/16 & 0.09 \\ \cline{3-4}\cline{7-8}
 &  & \multicolumn{2}{c|}{$\Delta A = (0.4 \pm 0.4) \%$} &  &  & \multicolumn{2}{c|}{$\Delta A = (0.8 \pm 0.6)\%$} &  &  \\ \hline
Misaligned correlation & Covariance & 5.7/11 & 0.89 & 8.3/16 & 0.94 & 5.8/8 & 0.67 & 15.8/16 & 0.47 \\ \hline
90\textendash 150 GHz null tests: & Foregrounds &  &  &  &  &  &  &  &  \\
- MV, map level diff. &  & 6.2/11 & 0.86 & 12.6/16 & 0.70 & 16.1/8 & 0.04 & 23.2/16 & 0.11 \\
- TT, map level diff. &  & 13.4/11 & 0.27 & 17.5/16 & 0.35 & 12.7/8 & 0.12 & 20.4/16 & 0.20 \\
- MV, band power diff. &  & 15.0/11 & 0.18 & 20.3/16 & 0.21 & 2.7/8 & 0.95 & 9.3/16 & 0.90 \\
- TT, band power diff. &  & 9.9/11 & 0.54 & 12.5/16 & 0.71 & 6.8/8 & 0.56 & 10.3/16 & 0.85 \\ \hline
\websky null test & Foregrounds & \multicolumn{2}{c|}{bias $< 0.02 \sigma$} &  \multicolumn{2}{c|}{} & \multicolumn{2}{c|}{bias $< 0.02 \sigma$} &  \multicolumn{2}{c|}{}   \\ \hline
Baseline vs CIB-depr. & Foregrounds & 11.5/11 & 0.40 & 13.57/16 & 0.63 & 14.7/8 & 0.07 & 18.3/16 & 0.30 \\\cline{3-4}\cline{7-8}
 &  & \multicolumn{2}{l|}{$\Delta A = 0.03 \pm 0.13$} & & & \multicolumn{2}{l|}{$\Delta A = 0.03 \pm 0.19$} & &\\ \hline
North vs South & Combined analysis & 8.6/11 & 0.66 & 15.3/16 & 0.50 & 2.4/8 & 0.97 & 7.5/16 & 0.96 \\
\hline
\end{tabular}%
}
\caption{Summary of the consistency and null tests conducted for estimating the angular cross-power spectrum between ACT DR6 CMB lensing and the SDSS BOSS galaxy samples. For each test, the primary purpose is listed and the test results are given for four different cases: for each of ACT $\times$ CMASS and ACT $\times$ LOWZ the results for the baseline cosmological analysis range and an extended validation range are listed including the $\chi^2$ value, the number of degrees of freedom (DOF), and the PTE value. For the signal recovery on the simulations test and the comparison of the baseline map to an alternative approach using CIB-deprojection, we list the amplitude difference constraints in the cosmological ranges. Finally, for the \websky null test, we constrain the residual bias to the analysis in each bin as a fraction of the measurement uncertainty $\sigma$. See \cref{sec_consistency_and_null_tests} for a discussion of each test.}
\label{table_null_tests}
\end{table*}

\subsubsection{Accurate input recovery} \label{sec_input_recovery}

We test if the analysis pipeline accurately recovers the input angular cross-power spectrum $\hat C_\ell^{\kappa g}$ for our fiducial cosmology. We use the 400 simulated ACT CMB lensing and galaxy map realizations discussed in \cref{sec_simulated_map_realizations} and apply our analysis pipeline to each combination. This test covers our full analysis pipeline including the normalization correction (see \cref{sec_norm_correction}) and also tests if the approximation of using the CMB mask squared as an approximation for the mask of the CMB lensing map is sufficiently accurate for the analysis.

We have tested that the mean angular cross-power spectra from all $N_{\rm sims} =400$ simulations recover the input signal. Here the input does not have an uncertainty, therefore the covariance in \cref{chi_sq_diff} is given by scaling the covariance of an individual measurement $\textrm{Cov}_{\rm mean} = \textrm{Cov}/N_{\rm sims}$. In the cosmological analysis ranges we find PTE values of $\textrm{PTE} = 0.86$ and $\textrm{PTE} = 0.32$ for ACT $\times$ CMASS and ACT $\times$ LOWZ and using the baseline simulation-based covariance. This shows that within statistical expectations we recover the input signal. We can also translate this into the absolute deviation from the input for our pipeline test. This depends on the number of simulations used. We fit the theory curve to the mean recovered signal with a free amplitude $A$, with perfect recovery yielding $A=1$. We find for $\Delta A = A -1 $ values of $\Delta A = (0.4 \pm 0.4) \%$ and $\Delta A = (0.8 \pm 0.6)\%$ for ACT $\times$ CMASS and ACT $\times$ LOWZ respectively. We conservatively account for the level at which we validated the amplitude in our systematic error budget. We use 95\% upper limits of $1.2\%$ and $2.0\%$ respectively which are small compared to the statistical uncertainties of the $E_G$ estimates.  

In \cref{table_null_tests} we list these values and additionally show the PTE values for our extended validation range where we also find that our pipeline accurately recovers the input.

\subsubsection{Misaligned correlation null test} \label{sec_misaligned_corr}

\begin{figure}
\includegraphics[width=\columnwidth]{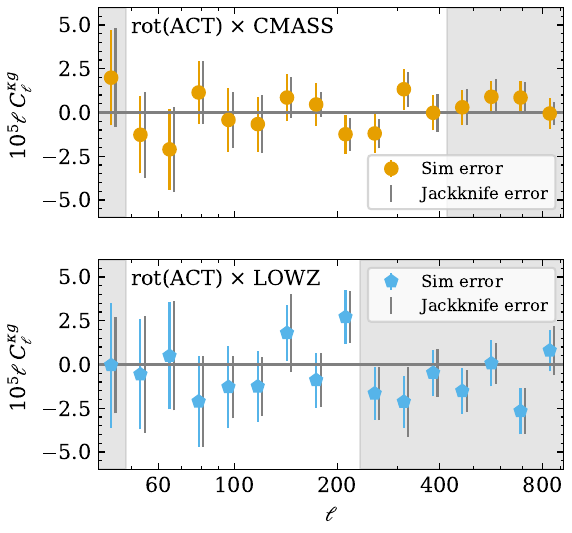}
\caption{Results for the misalignment null test of the angular cross power spectra following a rotation of the CMB map for [Upper] ACT $\times$ CMASS and [Lower] ACT $\times$ LOWZ with uncertainties based on simulations as a baseline [color] and jackknife errors [gray], slightly shifted to the right for clarity. 
See \cref{sec_misaligned_corr} for details.}
\label{fig:misaligntest}
\end{figure}

By misaligning the CMB lensing map and galaxy overdensity map purposefully, the cosmological signals become offset, and therefore we expect a null result for $\hat C_\ell^{\kappa g}$. By considering the PTE value of this null test we can test whether our covariance estimate is consistent with the data. If the PTE value is close to 1 this could indicate that the uncertainty is overestimated, i.e. the simulations we use to build our covariance have an excess variance that is not represented in the data. However, for a PTE value close to 0 the opposite would be the case: this would indicate the uncertainty of the measurement is underestimated and our simulations have less variance than the data. 

To perform this test we rotate the ACT CMB lensing map by 180 $\deg$ along the direction of RA. In this way, the overlap between the SDSS North and South patches is approximately switched. In \cref{fig:misaligntest} we show the resulting null angular cross-power spectra for the rotated ACT CMB lensing map cross-correlated with CMASS and LOWZ. We again find that the uncertainty estimated from performing this on our sets of 400 simulations is consistent with a data-based jackknife approach. Evaluating the PTE value of the null test we find PTE=0.89 for ACT $\times$ CMASS and PTE=0.67 for ACT $\times$ LOWZ in our cosmological analysis range. In the extended validation range, which includes the grayed-out scales shown, the PTE values are 0.94 and 0.47 respectively. The PTE values are all within statistical expectations. This gives further confidence beyond our direct comparison of different error techniques that our covariance estimate accurately captures the uncertainty in the data. 

\subsubsection{Websky foreground bias estimate} \label{sec_websky}

\begin{figure*}
\includegraphics[height=0.9\columnwidth]{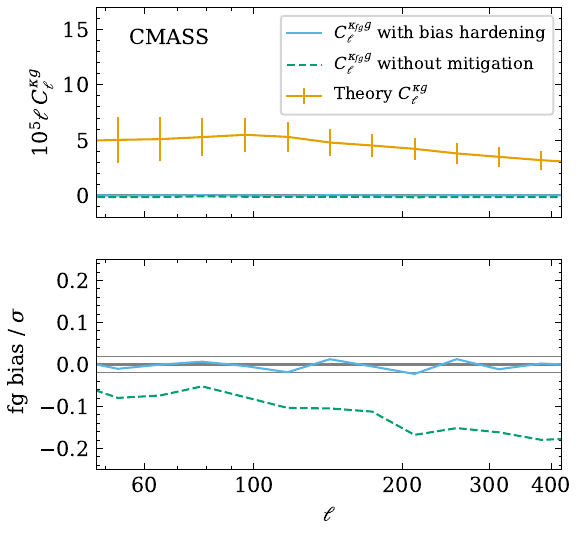}
\includegraphics[height=0.9\columnwidth]{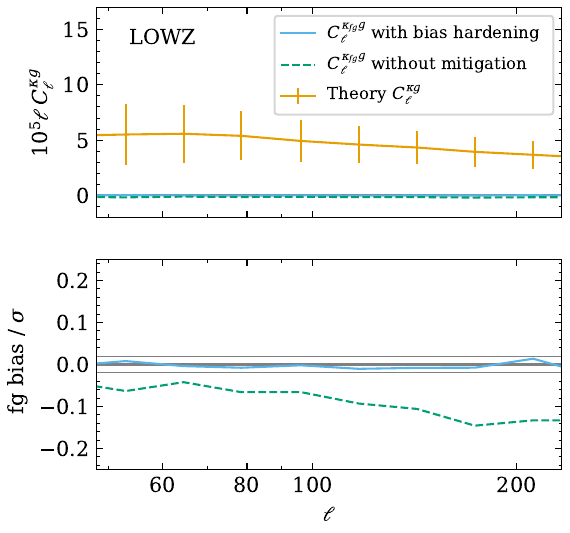}
\caption{We show that the lensing reconstruction approach of the ACT DR6 map that includes bias hardening sufficiently removes bias for the cross-correlation measurement from extragalactic foregrounds for the case of the \websky simulation. Shown is the null test of cross-correlating the $\kappa$ reconstruction performed on only the foreground contributions with the galaxy samples when using bias hardening [blue full line] and without mitigation [green dashed line]. Panels show the case of cross-correlation with [Left] a CMASS-like sample and [Right] for a LOWZ-like sample. [Upper] The panels show the null test compared to the expected cross-correlation signal [orange solid line] with errors showing the analytic uncertainty of the binned measurements [orange error bars]. [Lower] The estimated bias is shown as a fraction of the observational uncertainty. While there is a 0.1 - 0.2 $\sigma$ bias without mitigation, the bias-hardened estimator used for this analysis has a bias of less than $0.02 \sigma$ [gray horizontal lines] in the cosmological range. See \cref{sec_websky} for details. 
}
\label{fig:webskyforeground}
\end{figure*}

As the resolution of CMB measurements improves, additional analysis challenges come into focus. Of particular interest is the issue of extragalactic foregrounds that can affect the measured $\kappa$ map due to contamination in the measured CMB temperature and polarization \citep{vanEngelen2012,Osborne2014}. 

This is highly relevant for cross-correlation analyses as extragalactic foregrounds can also be correlated with the galaxy samples and thereby introduce spurious correlations that affect the measured signal. To mitigate the issue of foregrounds in the reconstruction of the ACT DR6 lensing map a multi-step approach was used. This includes explicit masking of known sources and for the reconstruction of the $\kappa$ map, a profile hardening technique.

The approach for the ACT DR6 lensing maps to mitigate bias from extragalactic foregrounds was extensively validated by \citet{MacCrann2023} for an auto-correlation analysis. The authors point out that while their work gives confidence to the auto-correlation analysis the impact can be different for cross-correlation analyses and it is therefore recommended to test the impact for the specific tracer one correlates with. Therefore in this work, we explicitly test the bias from extragalactic foregrounds for the cross-correlation with SDSS CMASS and LOWZ.

We test the foreground mitigation scheme on the \websky simulation \cite{Stein2019,Stein2020}. Specifically, we run the ACT reconstruction pipeline on the foreground-only CMB temperature map that models the contributions of CIB, kSZ, tSZ, and radio point sources to the observed CMB temperature anisotropy. All of these act as contaminants in the reconstruction of the CMB lensing signal. Applying the reconstruction pipeline on the foreground-only map gives the residual signal these foregrounds leave in the reconstructed $\kappa$ map. If these foreground residuals are correlated with a galaxy map constructed from the same simulation this would indicate that the foregrounds would contaminate our observed cross-correlation measurement. Therefore a null result builds confidence that the reconstruction approach sufficiently mitigates foreground contamination from extragalactic sources.

The \websky simulation includes a halo catalog that we can leverage to build galaxy samples that match the properties of our galaxy samples to test the mitigation for a comparable sample. We construct a CMASS-like galaxy catalog from the \websky halo catalog by applying an HOD which is tuned to closely match the CMASS galaxy sample \citep{Tinker2012,Beutler2014}, following previous approaches \citep[e.g.][]{White2014,Pullen2016}. The HOD has the form 
\begin{align} 
 \langle N_{\rm cen} \rangle_M &= \frac{1}{2} \left[ 1 + \textrm{erf} \left( \frac{\log M - \log M_{\rm min}}{\sigma}\right) \right] ,  \\
 \langle N_{\rm sat} \rangle_M &= \langle N_{\rm cen} \rangle_M \left( \frac{M}{M_{\rm sat}}\right)^\alpha \exp \left( \frac{- M_{\rm cut}}{M}\right),
\label{eq:HOD}
\end{align}
with best fit values $M_{\rm min} = \SI {9.319e12}{\Msun \h^{-1}}$, $\sigma = 0.2$, $M_{\rm sat} = \SI{6.729e13}{\Msun \h^{-1}}$, $\alpha = 1.1$ and $M_{\rm cut} = \SI{4.749e13}{\Msun \h^{-1}}$ \citep{White2014}.
For each halo, we sample observed galaxies based on $\langle N_{\rm cen} \rangle_M$ and place them in the center of the halo. For each halo that has a central galaxy, we Poisson sample satellite galaxies based on $\langle N_{\rm sat} \rangle_M / \langle N_{\rm cen}\rangle_M$ and placed in a random direction from the center with the radius sampled from a \citet{Navarro1996} (NFW) profile using halo concentration parameters from \citet{Duffy2008}. Finally, we subsample the galaxies to match the number density and redshift distribution of the CMASS sample. This results in a CMASS-like galaxy catalog which is correctly correlated with the other data products of the \websky simulation. For LOWZ we follow a similar approach implementing the HOD from \citet[Table 3, column Mean Full in][]{Parejko2013} with a slightly different functional form but otherwise proceed the same way to create a galaxy catalog. We also note that for the redshift range 0 to 1 the minimum halo mass captured by \websky is  $\geq \SI{1.2e12}{\Msun \h^{-1}} $ sufficient for our purposes for both samples. %

We confirmed the accuracy of our approach by correlating the galaxy catalog (based on \websky halos) with the \websky $\kappa$ map and checking that we recover a measurement of the cross-correlation that approximately matches the input cosmology for the simulation, in the cosmological range for our analysis.

In \cref{fig:webskyforeground} we show the result of our foreground test for CMASS and LOWZ. With the bias-hardened estimator used as a baseline for our analysis, there is a negligible bias of less than $0.02\sigma$ throughout the range in scales considered for our analysis for each cross-correlation. Without mitigation the resulting bias would be significant, reaching around $0.2\sigma$ in our cosmological range. Therefore we conclude that the bias hardening approach sufficiently reduces the bias from extragalactic foregrounds for the cross-correlation measurements with CMASS and LOWZ that we present in this work.

\subsubsection{90-150 GHz null test} \label{sec_90150_nulltests}

\begin{figure*}
\includegraphics[width=\columnwidth]{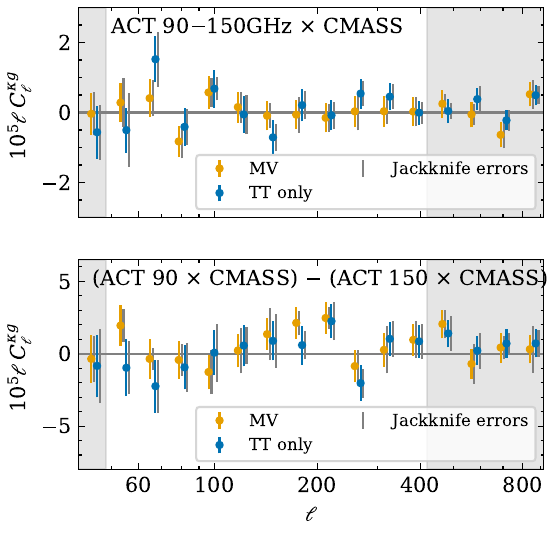}
\includegraphics[width=\columnwidth]{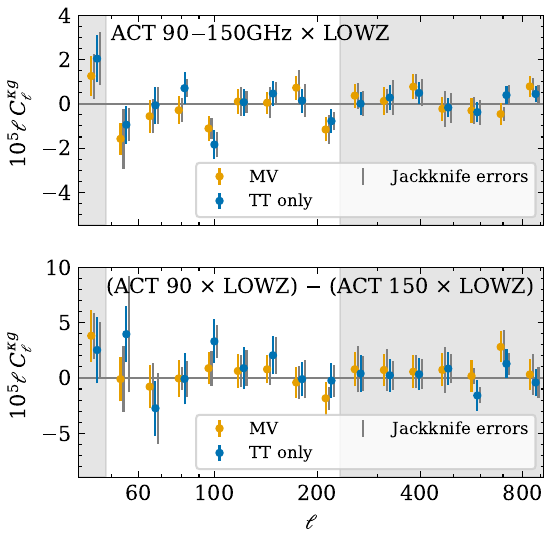}
\caption{Testing the angular cross-power spectrum estimates for contamination from foregrounds by nulling the cosmological signal in the ACT lensing map through differences of the $90~ \textrm{GHz}$ and $150~ \textrm{GHz}$ CMB observations. Shown are the correlations of the residual lensing maps without cosmological signal with CMASS [left] and LOWZ [right]. We show the difference taken at the map level [top panels] and bandpower level [bottom panels] and we show the result when using the baseline minimum variance combination of temperature and polarization information [MV] as well as temperature alone [TT only]. The baseline errors are based on our set of simulations with jackknife error estimates shown for comparison [gray]. See \cref{sec_90150_nulltests} for details.}
\label{fig_90150nulltest}
\end{figure*}

To test for contamination of our estimate from foregrounds we can use tests on the data itself in addition to the simulation-based test discussed in \cref{sec_websky}. One way is to purposely null the cosmological signal in the CMB lensing map and test if there is any residual correlation with the galaxy maps which would indicate some foreground contamination. 

By taking the difference between the $90~\textrm{GHz}$ and $150~\textrm{GHz}$ maps we null the CMB lensing information in the primary CMB. This also nulls frequency-independent components like kSZ but other foreground effects can still be in the map. By running the lensing reconstruction pipeline on the difference we create a null map that should be uncorrelated with the galaxy maps which we can then test. We can also run the reconstruction on the individual frequency maps and then take the difference in the final bandpowers. Finally, we can limit this test to the temperature data alone and check if there is a difference compared to the baseline minimum variance combination of temperature and polarization information. 

In \cref{fig_90150nulltest} we show the resulting angular power spectra for all these variations of the 90\textendash 150~GHz null tests for both ACT $\times$ CMASS and ACT $\times$ LOWZ. In \cref{table_null_tests} we list the PTE values for the set of null tests. All but one of the 90\textendash 150~GHz null tests have PTE values between 0.05 and 0.95.

The PTE value for the 90\textendash 150~GHz null test at the map level using both temperature and polarization and for ACT $\times$ LOWZ is marginally below 0.05 with a value of PTE=0.04. Given the number of null tests performed in this analysis having a PTE value this low is not surprising but it motivates a check for those results to identify if there is a concern. The same test for only the temperature information passes and the test at the bandpower level also passes further indicating this could be a statistical outlier. We find this null test does not fail if we add or remove a bin at the small-scale end. When adding a bin at the large-scale end the test still marginally fails, when removing one it passes. The test does pass for our full extended validation range. This dependence on the test range could be an indication that this is indeed a chance failure. In the plot (orange points in the top right panel of \cref{fig_90150nulltest}), we do not see clear outliers and there is no scale-dependent trend visible. Overall we conclude that this PTE value of 0.04 is within statistical expectation and we did not identify a reason for concern. In \cref{sec_summary_of_PTE} we summarize all PTE value tests in this analysis and evaluate whether there are more low PTE values as expected for the number of tests performed. 
The difference map does not show a statistically significant correlation with the
galaxy maps.

\subsubsection{Consistency with CIB-deprojection approach} 
\label{sec_CIB_deprojected}

As part of the DR6 lensing analysis, a different foreground mitigation strategy using CIB-deprojection was considered instead of the profile hardened estimator used for the baseline map \citep{Qu2023,MacCrann2023}. As they have distinct sensitivity to potential foreground contamination, testing the two approaches for consistency in terms of the results for the angular cross-power spectrum is another test for potential foreground contamination as well as for the robustness of the analysis.

We compare the resulting angular cross-power spectra with this alternative CIB-deprojected map with the baseline map and test for consistency. The CIB-deprojected map uses additional external information from \textsl{Planck} and has a restricted mask called the HILC mask. For a consistent comparison, we limit the baseline mask to the same footprint and recalculate the angular cross-power spectrum for this comparison. 

The two maps are significantly correlated as they are based on the same CMB observations. Therefore, when testing them for consistency, we account for the cross-covariance between the two angular power spectrum measurements. We here leverage that the simulated map realizations for both the baseline and CIB-deprojected maps are based on the same CMB lensing realizations and therefore correctly correlated. We can estimate the covariance between the measurements with \cref{eq_cov_sim_based}. We find that the two measurements in the same bins are $(98-99) \%$ correlated. 

We calculated the angular cross-power spectra for the CIB-deprojected ACT CMB lensing map and the baseline map restricted to the same footprint. We found that the results closely match. We test for consistency between the two measurements using a PTE value test, accounting for the correlation. For the cosmological analysis range, we find PTE values of 0.40 and 0.07 for ACT $\times$ CMASS and ACT $\times$ LOWZ respectively. In the extended validation range, we find PTE values of 0.63 and 0.30 respectively. All PTE values indicate consistency between the two foreground mitigation approaches.

\subsubsection{Consistency between North and South observations} \label{sec_north_v_south}

\begin{figure}
\includegraphics[width=\columnwidth]{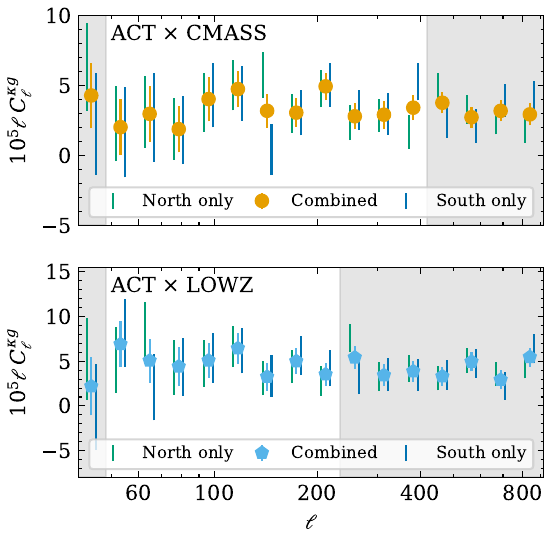}
\caption{We compare the baseline angular cross-power spectra of the ACT CMB lensing map with the full CMASS [top panel] and LOWZ [bottom panel] datasets with results using only the North or South data. For each, we show the baseline result [orange/light blue middle points] with the north-only result [left green] and south-only result [right blue].}
\label{fig:NStest}
\end{figure}

SDSS has two separate patches on the sky, in the North and the South. The two patches are often observable in different seasons and include different latitudes. Higher latitudes can have less foreground contamination and the patches could have different calibrations, seasonal variations, and selection functions \cite{Chen2022}. In this analysis we proceed analogous to the approach for the auto-correlation developed in \cite{Wenzl2024_EGestimator}, to test if, for the cross-correlations measured in this work, there are any statistically significant differences between the patches that might indicate that systematic effects are relevant.  If no such differences are found it motivates an analysis of the data as a combined whole rather than two distinct patches. 

The angular power spectra are calculated using the full CMASS and LOWZ maps. We check the measurements for consistency between the North and South patches of the galaxy observations (see \cref{fig:masks}). We find that both North and South patch correlations agree statistically. 

In \cref{fig:NStest} we show the result for the angular cross-power spectra calculated for the North and South patches separately. We evaluate the consistency between the two measurements via a PTE value test. Since the two angular cross-power spectra do not overlap in the area and our largest scales considered ($\ell = 48$) are significantly smaller than the separation between the two patches we assume them to be uncorrelated. In the cosmological analysis range, we find PTE values of 0.66 and 0.97 for ACT $\times$ CMASS and ACT $\times$ LOWZ respectively. For the extended range, the PTE values are 0.50 and 0.96 respectively. In all cases, the PTE values indicate full consistency between the North and South patches. We note that the values for ACT $\times$ LOWZ are marginally above 0.95 which indicates that they agree closer than expected in 95\% of tests. Since we perform a large number of PTE tests we assume that this is a chance occurrence and evaluate if there is a higher-than-expected number of high PTE values among our PTE tests when summarizing the distribution of all of them in \cref{sec_summary_of_PTE}.

Overall we conclude that the analysis is consistent between the North and South maps. Any difference in the effective galaxy bias for the two patches, for example, caused by differences in the selection of the galaxies, is not large enough to be statistically relevant for the uncertainties in our angular cross-power spectra.

\subsubsection{Summary of PTE statistics} \label{sec_summary_of_PTE}

\begin{figure}
\includegraphics[width=\columnwidth]{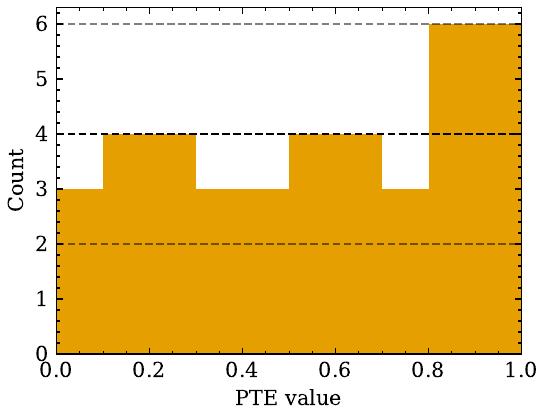}
\caption{We report a total of 40 PTE value tests and here show a histogram of their distribution to check if they follow a uniform distribution within statistical expectations. The Poisson expectation for 10 bins shown is $4.0 \pm 2.0$ per bin [black dashed lines]. Therefore the distribution is consistent with the expected flat PTE distribution for an unbiased statistical analysis. Note that some of the reported PTE values are strongly correlated which is not considered here.  }
\label{fig_pte_distr}
\end{figure}

Throughout this work, we report a large number of PTE tests. If a measurement is consistent with the null hypothesis then the PTE value by definition should be a sample from a uniform distribution between 0 and 1. When reporting many PTE tests the probability of detecting outliers increases. Under the assumption that the distinct reported PTE values are independent, we can analyze whether together they follow this uniform distribution. If they do it indicates that the statistical tests performed are unbiased and show consistency between the measurement and our null hypothesis. If they are not consistent it could point to issues with the test, the analysis pipeline, or an inconsistency of the measurement with GR. We here analyze both the overall distribution and check for stronger outliers that would be expected for the number of tests we perform.

We report a total of 40 tests. For an unbiased set of tests, each tenth between 0 and 1 should have a number of PTE values sampled from a Poisson distribution with a mean of 4 with a statistical variation of 2. If our distribution deviates from this it could indicate that the PTE value estimates are biased due to, for example, under- or overestimation of the uncertainty, or, for the cosmological tests, due to a disagreement with the GR-based predictions. We show a histogram of the PTE tests in bins of 10\% in \cref{fig_pte_distr}. For each bin in the histogram, the number of PTE values is within statistical expectations.  

For an individual test, a PTE value of less than 0.05 is generally considered statistically significant as it has a change of only 5\% to occur by chance (approximately the equivalent of 2$\sigma$ for the case of a Gaussian). However, when reporting multiple tests some outliers become more probable. For 40 tests having one outlier PTE value at or below 0.05 by chance has a probability of approximately 87\% and having an outlier at or below 0.01 has a chance of approximately 33\%. Among the reported PTE tests there are no outliers with PTE values below 1\% or above 0.99\%. The lowest PTE value across all PTE tests is 0.04 for a null test discussed in \cref{sec_90150_nulltests}. Therefore we conclude that the reported PTE values are fully consistent with statistical expectations indicating that our suite of null and consistency tests confirms the accuracy of our analysis pipeline.

\subsection{Measurement results for cross-correlation of ACT and SDSS BOSS} \label{sec_measurement_angular_cross_power_ACT_SDSS}

\begin{figure*}
\includegraphics[width=1.02\columnwidth]{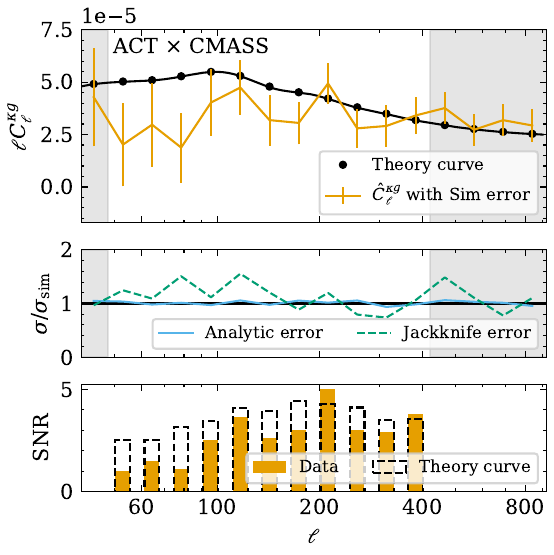}
\includegraphics[width=0.98\columnwidth]{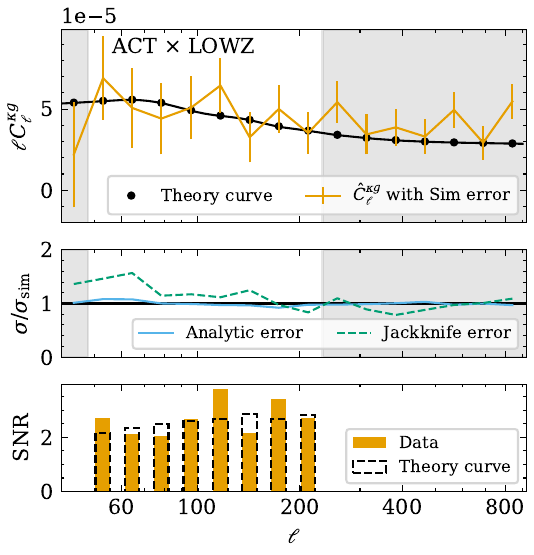}
\caption{Measurment results for the angular cross-power spectrum between the ACT DR6 lensing map and CMASS [left] and LOWZ [right]. [Upper] The measurement result [orange error bar plot] and the theory curve for the fiducial cosmology [black line]. Note that the fiducial cosmology has a free amplitude, and the theory curves have not been fit to the cross-correlation measurements. Shown is the full extended validation range with the scales outside the cosmological range grayed out. The middle panel shows a comparison of the baseline simulation-based marginalized uncertainty compared with an analytic estimate [blue line] and a jackknife estimate [dashed green line]. [Lower] shows the signal-to-noise ratio [orange filled] and, for better comparison between surveys, the ratio of the theory curve to the noise [gray dashed].  }
\label{fig:cross_corr_results}
\end{figure*}

We report the results of applying the analysis approach to measure the angular cross-power spectrum between the ACT DR6 CMB lensing map and the SDSS BOSS CMASS and LOWZ galaxy samples. The analysis pipeline, which has been validated through an extensive set of null and consistency checks, results in the angular cross-power spectra measurements presented in \cref{fig:cross_corr_results}.

We show the measurements using ACT together with CMASS and LOWZ respectively. The top panels show the measurements with the baseline simulation-based error estimates. The white background range indicates the scales used for the cosmological analysis. The full range plotted is the extended validation range for which we also performed null and consistency tests. For comparison, the theory curve for our fiducial cosmology is also shown with dots indicating the binned measurement using the \namaster coupling matrix. Note that this theory curve is not tuned to the amplitude of the measurement: the galaxy bias used is based on the amplitude of the angular auto-power spectrum of the galaxy data. In this work, we focus on constraining the $E_G$ statistic for which we directly use the measurements as part of the estimator, and the predicted theory values for $E_G$ are estimated from independent constraints on $\Omega_{m, 0}$. In future work, the cross-correlation measurements we make available as part of this work can be consistently fitted with angular power spectra theory curves by varying the galaxy bias and all cosmological parameters. Specifically, we can visually see that for ACT $\times$ CMASS the theory curve tuned to the measured galaxy auto-correlation amplitude appears slightly high compared to the cross-correlation measurement. This implies that when jointly fitting the cross-correlation one will likely find a lower best fit $\sigma_8$ value, which quantifies the amplitude, than for the CMASS galaxy data alone. To evaluate whether this difference is significant or within statistical expectations careful theoretical modeling of the angular power spectra is needed in future work. We here provide the tested and validated cross-correlation measurements between ACT and CMASS that could be used for such an investigation. The ACT DR6 lensing data has already been analyzed in this way for combinations with galaxy data from unWISE \citep{Farren2023} and similar investigations have been done for \textsl{Planck} lensing data \citep[e.g.][]{Chen2022}. For this work, we focus on interpreting the measurement at linear scales through the $E_G$ statistic to interpret implications about gravity. 

In the middle panels, the error for each bin is compared. The simulation-based estimate of the errors is closely consistent with the analytic estimate. This consistency also extends to the off-diagonal elements of the covariance as discussed in \cref{sec_covariance_estimation}. We also show the uncertainties based on a jackknife approach which have larger statistical uncertainty and are a systematic overestimate. The jackknife errors are consistent with the other estimates within statistical expectations. 

Finally, in the bottom panels, we show the signal-to-noise ratios (SNR) of the measurements. The values based on dividing the signal by the error are shown in solid orange. In addition, we also show the theory value for the fiducial cosmology divided by the error which is more comparable between surveys for low SNR measurements. We can also calculate the SNR for all bandpowers in the cosmological range combined as
\begin{align}
    \textrm{SNR} = \sqrt{ \sum_{\ell \ell'} \hat C_\ell^{\kappa g} \textrm{Cov}_{\ell \ell'}^{-1}\hat C_{\ell'}^{\kappa g}}.
\end{align}
We find a total SNR in our cosmological range of 10.0 for ACT $\times$ CMASS and 8.5 for ACT $\times$ LOWZ. When using the theory curve of the fiducial cosmology instead of the measurement for comparability between different surveys the expected total SNR is 12.4 for ACT $\times$ CMASS and 7.9 for ACT $\times$ LOWZ. For both SNR estimates we here use the simulation-based covariance. 

Based on the measured angular cross-power spectra we derive constraints on the $E_G$ statistic for the ACT DR6 CMB lensing data combined with CMASS and LOWZ in \cref{sec_EG_measurement}. In the following, we combine the measurement with results based on \textit{Planck} CMB lensing data.

\subsection{Combined cross-correlation measurement with ACT and Planck} \label{sec_combined_cross_correlation}

\begin{figure}
\includegraphics[width=\columnwidth]{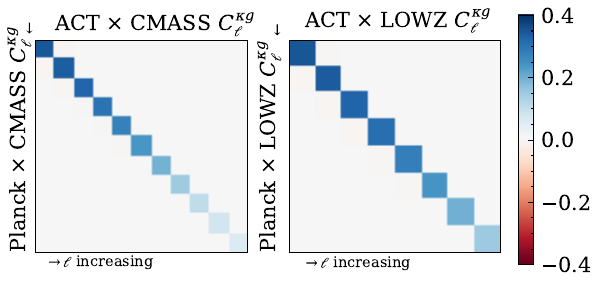}
\caption{The correlation coefficients for the analytic cross-covariance between the $C_\ell^{\kappa g}$ measured with ACT $\times$ CMASS/LOWZ [right] and \textit{Planck} $\times$ CMASS/LOWZ [left].}
\label{fig_cross_covariance}
\end{figure}

While the ACT DR6 lensing map has a higher angular resolution than the \textit{Planck} lensing map, the latter covers a larger part of the sky. In particular, the \textit{Planck} data covers the SDSS BOSS samples almost completely, covering approximately $96\%$ of each. The ACT DR6 lensing map covers $34\%$ of the CMASS map and $37\%$ of the LOWZ map. Because of these distinct strengths, the datasets are highly complementary and one can maximize the overall constraining power by combining the angular cross-power spectra from ACT $\times$ SDSS and \textit{Planck} $\times$ SDSS. For \textsl{Planck} $\times$ SDSS we can use the $C_\ell^{\kappa g, Planck}$ measurements presented in \citet{Wenzl2024_EGestimator} which have been consistently estimated with the same binning scheme and the same processing of the SDSS BOSS datasets.

We take the minimum variance-weighted combination of the two angular cross-power spectra. The details of the approach are described in \cref{combining_cross_corr_measurements}. We account for the correlation between the two measurements in the overlap region. Correlation between the two measurements comes from the overlap region where we observe the same cosmic realizations of the CMB and the same galaxy maps between the two measurements. The measurement noise in the primary CMB map can be assumed to be independent between the two surveys. 

We estimate the cross-covariance analytically with \namaster and use the analytic covariance estimates throughout the combined measurement to avoid numerical issues from the uncertainty in the simulation-based covariance. We show the cross-covariances in \cref{fig_cross_covariance}. Plotted is the correlation given as $\textrm{CrossCov}_{\ell \ell'} / (\sigma_{C_\ell^{\kappa g}} \sigma_{C_{\ell'}^{\kappa g, Planck}})$. 
On the largest scales in our cosmological analysis range the correlation between the two measurements is up to 33\% for CMASS and 34\% for LOWZ.

We can discuss the improvement for the combined analysis in the context of the overall signal-to-noise ratio of the measurements for the example of CMB lensing $\times$ CMASS. We know that both measure the same underlying signal. To make a fair comparison of the constraining power of the two measurements one should compare them with fixed signal and the covariance from each measurement. We do this by using the theory curve for our fiducial cosmology and the analytic covariance for each measurement. For the analytic covariance, the expected SNR for the ACT $\times$ CMASS measurement is 12.3\footnote{We here for consistency use the analytic covariance. In the previous section, we discussed the SNR based on the simulation-based covariance which rounds to 12.4. } and for \textsl{Planck} $\times$ CMASS is 16.0. If the two measurements were uncorrelated the combined measurement would have an SNR ratio of 20.2. After accounting for the correlation between the datasets the minimum variance combination has an expected total SNR of 18.5. In the overlap region between ACT and \textsl{Planck}, the significantly higher fidelity in the ACT map means that the underlying \textsl{Planck} map does not contribute significant extra information. When removing the ACT DR6 region from the \textsl{Planck} map and using this restricted map for the cross-correlation the angular cross-power spectrum becomes mostly independent of the ACT-based measurement giving a combined expected SNR of 17.9. This illustrates how the two datasets are complementary: where we have ACT data available it dominates the signal due to its higher angular resolution but the \textsl{Planck} data has a larger area coverage resulting in a larger expected SNR for the combined measurement than either individually.

\begin{figure*}
\includegraphics[width=0.99\columnwidth]{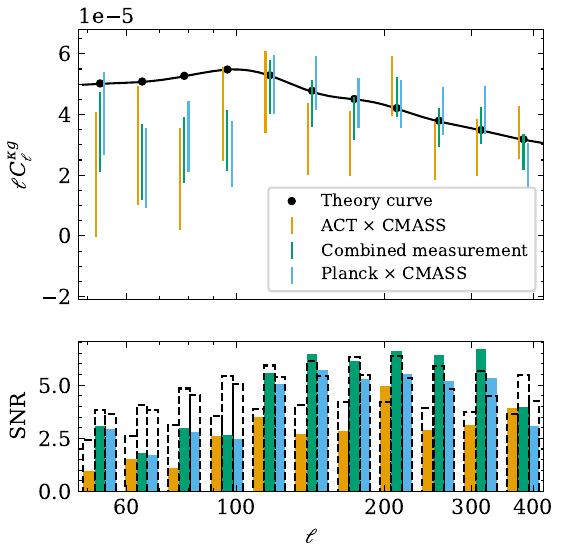}
\includegraphics[width=1.01\columnwidth]{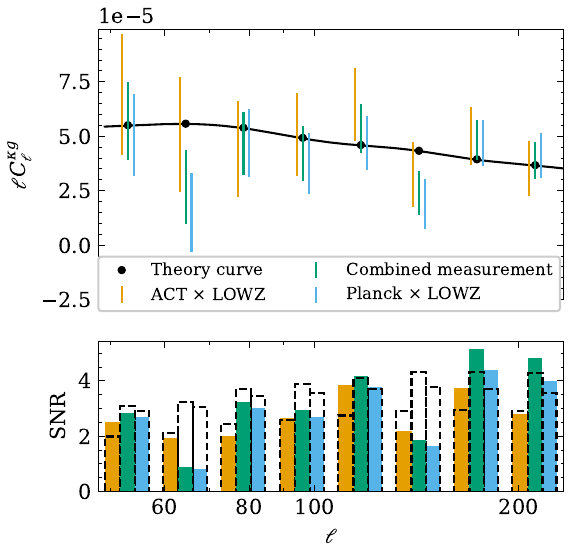}
\caption{Combined angular cross-power spectra for CMB lensing $\times$ CMASS [left] and CMB lensing $\times$ LOWZ [right]. For each, we show the ACT-based measurement [orange, left], the \textsl{Planck} based measurement [blue, right], and the minimum variance combination of the two [Green, middle], all using the analytic covariance. [Upper] The measurements and the theory curve for the fiducial cosmology (not fitted to the data). [Lower] The SNR [solid bars] and the SNR expected based on the theory curve and measurement covariance [gray, dashed]. For further details see \cref{sec_combined_cross_correlation}. }
\label{fig_combined_cross_corr}
\end{figure*}

In \cref{fig_combined_cross_corr} we show the results for combining the full baseline ACT $\times$ SDSS and \textsl{Planck} $\times$ SDSS angular cross-power spectra. We show the correlation with the CMASS sample on the left and with the LOWZ sample on the right. For each, we show the results using ACT CMB lensing discussed in \cref{sec_measurement_angular_cross_power_ACT_SDSS} and the results using \textsl{Planck} $\times$ SDSS presented in \citet{Wenzl2024_EGestimator}. We can visually see that the two measurements are closely consistent with each other, especially on large scales, which is expected since they are strongly correlated. Finally, we show the results for the minimum variance combination of the two measurements. In the bottom panels, we show the SNR for each measurement as well as the combination. We again additionally show the SNR estimates using the theory curves (dashed black lines) which represent a fair comparison between the measurements.

\section{\texorpdfstring{E\textsubscript{G}}{EG} measurement} \label{sec_EG_measurement}

\subsection{Measurement approach for \texorpdfstring{E\textsubscript{G}}{EG}}

Based on the measurements of the angular cross-power spectra between ACT CMB lensing and the SDSS BOSS CMASS and LOWZ samples we derive new constraints on the $E_G$ statistic through the estimator given in \cref{E_G_estimator}. We use the already calculated and characterized angular auto-power spectra $C_\ell^{gg}$ and RSD parameter $\beta$ for the LOWZ and CMASS samples presented in \citet{Wenzl2024_EGestimator}. The RSD analysis uses the published SDSS BOSS measurements \citep{GilMarin2016} and up-to-date theory modeling using the \velocileptors code\footnote{\url{https://github.com/sfschen/velocileptors}} \citep{Chen2020,Chen2021}. Note we use a plus in our notation to indicate when the auto-correlation and RSD measurements are included and a $\times$ when only the cross-correlation is considered. 

In \cref{sec_cross_covariance} we discuss the creation of correctly correlated simulated maps and the estimate of the cross-covariance between $C_\ell^{\kappa g}$ and $C_\ell^{gg}$.

We use estimates of $C_\ell^{gg}$ and $\beta$ for the full galaxy maps. To quantify the potential systematic caused by differences in galaxy bias between the full sample and the ACT DR6 overlap we constrain the galaxy bias difference and derive an estimate for our systematic error budget in \cref{sec_galaxy_bias_diff}. 

When calculating the $\hat E_G$ estimate using \cref{E_G_estimator} we account for the full covariance of $\hat C_\ell^{\kappa g}$ and $\hat C_\ell^{gg}$, including their cross-covariance (see \cref{sec_cross_covariance}). For this, we calculate the ratio distribution\footnote{The probability distribution for the ratio of variables from two known distributions.} of $\Gamma_\ell \hat C_\ell^{\kappa g} / \hat C_\ell^{gg}$. Then in a second step, we calculate the ratio distribution of this estimate and $\beta$ resulting in a probability density function (PDF) for $\hat E_G$ marginalized across the cosmological range or for individual bins $\hat E_G^{\ell}$. Notably, while we assume the measurements of $\hat C_\ell^{\kappa g}$ and $\hat C_\ell^{gg}$ to be Gaussian distributed we do not assume the ratio to be Gaussian as the PDF is expected to be significantly asymmetric \citep{Wenzl2024_EGestimator}. We report the median and the central $68.27\%$ confidence range of the PDF which generalize the best fit and $1\sigma$ uncertainty for the Gaussian case. For a full description of this calculation and a derivation of the PTE value tests to evaluate the results for consistency with scale independence ($\textrm{PTE}_{\operatorname{scale-indep}}$) and for consistency with a predicted value ($\textrm{PTE}_{\rm GR}$) see \citet{Wenzl2024_EGestimator}.

\subsubsection{Cross-covariance estimation} \label{sec_cross_covariance}

The angular cross-power spectrum between CMB lensing and galaxy clustering is significantly correlated with the angular auto-power spectrum of the galaxies. We estimate the cross-covariance between the measurements and account for this correlation in our calculation of $E_G$.

For our baseline result using ACT and SDSS BOSS, we leverage the set of 400 simulated map realizations to estimate the full covariance. As described in \cref{sec_simulated_map_realizations}, we have 400 simulated map realizations with the CMB lensing map and two galaxy maps for each with one galaxy map for the angular auto-power spectrum and the other for the angular cross-power spectrum where an additional redshift dependent weight was applied. For each realization, the three maps are correlated as expected for our fiducial cosmology and they contain the same noise level as our measurements. Using \cref{eq_cov_sim_based} we estimate the simulation-based covariances of $C_\ell^{\kappa g}$ and $C_\ell^{gg}$ individually as well as the cross-covariance between them. For the full analytic covariance, we use \namaster as described in \cref{sec_simulated_map_realizations}. For the covariance of $C_\ell^{\kappa g}$ we use a shot noise of $N^{\rm shot}_{\rm cross}$ and for $C_\ell^{gg}$ we use $N^{\rm shot}_{\rm auto}$ as given by \cref{eq:shotnoise}. For the cross-covariance, we use $\sqrt{N^{\rm shot}_{\rm cross} N^{\rm shot}_{\rm auto}}$. We calculated the full simulation-based and analytic covariances. We find consistent results between the two methods within the statistical uncertainty of the simulation-based covariance. For the same bin, the correlation between the two measurements reaches around 33\% for ACT + CMASS and 28\% for ACT + LOWZ and drops off for smaller scales.

\subsubsection{Effective galaxy bias difference in ACT overlap}
\label{sec_galaxy_bias_diff}

The ACT DR6 lensing dataset does not cover the full SDSS CMASS or LOWZ footprints: The ACT DR6 lensing map covers 34\% of the CMASS galaxy map and 37\% of the LOWZ map. While the galaxy samples are designed to be uniform across their footprint, residual area dependence could remain. Differences in the galaxy sample in the overlap region compared to the full footprint can lead to differences in the effective linear galaxy bias $b$ of the sample. If this difference is statistically significant compared to our measurement uncertainty it could bias the $E_G$ result because the estimator relies on cancellation of the galaxy bias between the measurements. To investigate this we quantify the difference in effective galaxy bias between the ACT overlap and the full sample to evaluate if it is statistically relevant and add a limit to our systematic error estimate.

We can perform a test for the ratio of the inferred $C_\ell^{gg}$ between the overlap and the full sample. Assuming the underlying matter power spectra and redshift distribution are sufficiently similar for both the full sample and the overlap with ACT, this ratio directly tests if the effective linear galaxy bias is consistent between the two. By assuming that a constant linear bias describes the data well in our cosmological range, and therefore $C_\ell^{gg}\propto b^2$, %
we can put a constraint on the fractional difference in bias between the two $\Delta b / b$. %
Since this approach avoids having to marginalize over uncertainty in the matter power spectrum this gives us a strong constraint on the fractional bias difference of the bias between the samples.

\begin{figure}
\includegraphics[width=\columnwidth]{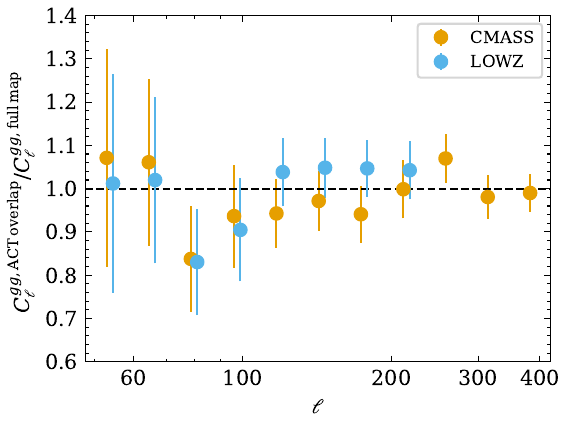}
\caption{We test for a difference in the galaxy auto-correlation between the full galaxy samples and the overlap region with ACT. Shown is the ratio of $C_\ell^{gg}$ in the overlap region to the full sample for CMASS [orange] and LOWZ [blue]. The errors shown are the uncertainty on the ratio under consideration of the correlation between the two measurements. For each case, only the bins in the cosmological analysis range are shown. See \cref{sec_galaxy_bias_diff} for further details.}
\label{fig_bias_overlap_test}
\end{figure}

In \cref{fig_bias_overlap_test} we show the ratio of $C_\ell^{gg, \rm ACT\, overlap}$ in the overlapping region with ACT DR6 and $C_\ell^{gg, \rm full\, map}$ of the full galaxy samples for both CMASS in orange and LOWZ in blue. In each case, we show the bins in the cosmological analysis range. We note that we only constrain the difference, the absolute value is strongly degenerate with cosmological parameters. Since the two $C_\ell^{gg}$ measurements are based on overlapping data it is crucial to include the cross-covariance between the two. We use our set of simulations to estimate the full covariance including the cross-covariance. When assuming Gaussianity the covariance for the ratio of two correlated measurements $A_\ell, B_\ell$, so that $C_\ell = A_\ell / B_\ell$  is given by 
\begin{align}
&\textrm{CovN}(C_\ell, C_{\ell'}) = \nonumber\\ 
&\textrm{CovN}(A_\ell, A_{\ell'})  + \textrm{CovN}(B_\ell, B_{\ell'}) \nonumber\\ 
&- \textrm{CovN}(A_\ell, B_{\ell'}) - \textrm{CovN}(B_{\ell}, A_{\ell'}), \label{eq_cov_ratio}
\end{align}
where we have defined 
\begin{align}
    \textrm{CovN}(A_\ell, B_{\ell'}) \equiv \frac{\textrm{Cov}(A_\ell, B_{\ell'}) }{ A_\ell B_{\ell'}}.
\end{align} 
Then we fit a constant amplitude $A_{\rm Ratio}$ to the ratio accounting for the covariance of the ratio given by \cref{eq_cov_ratio}. We can translate this to a constraint on the fractional bias as
\begin{align}
    \frac{\Delta b}{b} = \sqrt{A_{\rm Ratio}} -1
\end{align}
For CMASS we get $\Delta b/b = (-0.9 \pm 0.9)\%$ and for LOWZ we get $\Delta b/b = (0.8 \pm 1.3)\%$. We do not find a statistically detectable difference in the galaxy bias between the full sample and the ACT overlap, indicating that the difference is small compared to our measurement uncertainty. We conservatively use the constraints to put limits on the systematic offset in the effective linear galaxy bias between the full sample and the overlap which we account for in our systematic error budget. We find
\begin{align}
    \left| \frac{\Delta b}{b_{\rm full}}\right| &< 2.7\% \, (95\% \textrm{ c.l.; CMASS}),  \\
    \left| \frac{\Delta b}{b_{\rm full}}\right| &< 3.4\% \, (95\% \textrm{ c.l.; LOWZ}).
\end{align}

For our baseline results, we use an RSD analysis and $C_\ell^{gg}$ measured for the full SDSS maps with $\beta C_\ell^{gg} \propto b$ while our $C_\ell^{\kappa g} \propto b$ measurement is restricted to the overlap with ACT. A difference in the effective bias directly translates into a systematic error for the $\hat E_G$ estimate: $\frac{\sigma_{\rm sys, eff. bias} (E_G)}{\hat E_G} = \frac{\Delta b}{b}$. The 95\% upper limits on the systematic based on $C_\ell^{gg}$ measurements are small compared to the statistical uncertainty of the $E_G$ measurements of $~18\%$ for ACT + CMASS and $~26\%$ for ACT + LOWZ.

The uniformity of the effective galaxy bias of the SDSS samples across the footprint permits us to combine our cross-correlation with the full auto-correlation and RSD analysis for the SDSS datasets \citep{Wenzl2024_EGestimator} to measure $E_G$.  For completeness, we also present the $E_G$ result when the auto-correlation is limited to the region overlapping with ACT (see \cref{sec_analysis_choices}).

\subsection{Baseline \texorpdfstring{E\textsubscript{G}}{EG} results}
\label{sec_EGresults}

\begin{figure}
\includegraphics[width=\columnwidth]{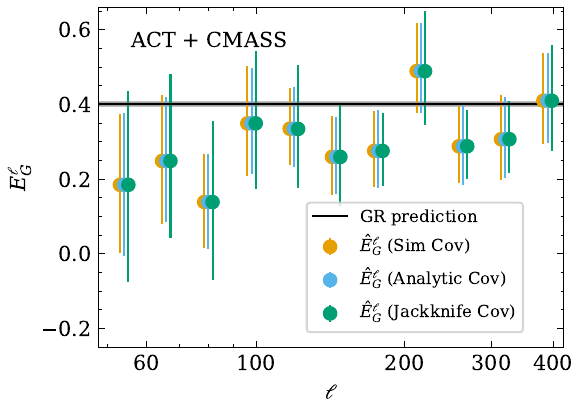}
\includegraphics[width=\columnwidth]{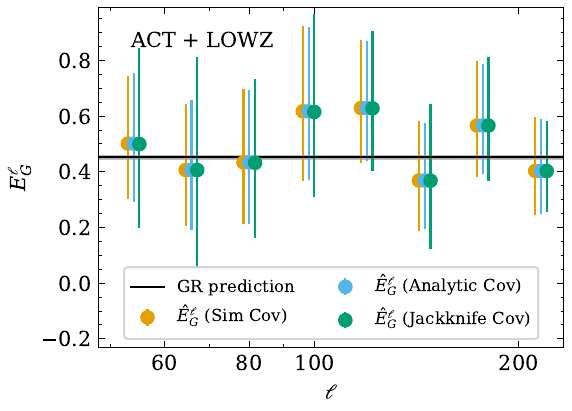}
\caption{Measurement of $E_G^{\ell}$ as a function of scale. Shown is $E_G^{\ell}$ measured based on the angular cross-power spectra $C_\ell^{\kappa g}$ measured in this work and external constraints on $C_\ell^{gg}$ and $\beta$ for the galaxy samples. [Upper] Constraints using the CMASS galaxy sample and the [Lower] using the LOWZ galaxy samples. For each case, the constraints using three different uncertainty estimates are shown: the baseline simulation-based covariance [orange], the analytic covariance [blue], and the jackknife covariance [green]. The ranges of scales shown are the cosmological analysis range for each case and the measurements are shifted slightly left and right for visibility. }
\label{fig_EG_ell_measurement}
\end{figure}

\begin{table*}%
\begin{tabular}{|l|l|l|l|l|l|}
\hline
 & $z_{\rm eff}$ & GR prediction $E_G^{\rm GR}$ & Measurement $\hat E_G$ & $\textrm{PTE}_{\operatorname{scale-indep}}$ & $\textrm{PTE}_{\rm GR}$ \\ \hline
ACT + CMASS & \multirow{2}{*}{$ 0.555$} & \multirow{2}{*}{$0.401\pm 0.005$} & $0.31^{+0.06}_{-0.05} (\textrm{68\% CI}) $ & 0.74 & 0.13 \\
ACT + \textsl{Planck} + CMASS &  &  & $0.34^{+0.05}_{-0.05} (\textrm{68\% CI}) $  & 0.53 & 0.31 \\ \hline
ACT + LOWZ & \multirow{2}{*}{$ 0.316$} & \multirow{2}{*}{$0.452\pm0.005$} & $0.49^{+0.14}_{-0.11} (\textrm{68\% CI}) $ & 0.95 & 0.71 \\
ACT + \textsl{Planck} + LOWZ &  &  & $0.43^{+0.11}_{-0.09} (\textrm{68\% CI}) $  & 0.46 & 0.86 \\
\hline
\end{tabular}
\caption{Overview of the $\hat E_G$ measurements and statistical tests presented in this work. Shown are the results for combining ACT DR6 CMB lensing and the SDSS BOSS CMASS and LOWZ samples and additionally combined constraints with \textsl{Planck} PR4 CMB lensing information. For each case the effective redshift of the measurement and the predicted value for four fiducial cosmology assuming $\Lambda$CDM are shown. The measurements are listed together with the uncertainties based on the $68.27 \%$ confidence intervals. For each measurement PTE value tests for the consistency of the measurement with scale independence as expected for GR ($\textrm{PTE}_{\operatorname{scale-indep}}$) and for consistency with the predicted value ($\textrm{PTE}_{\rm GR}$) are shown.}
\label{tab_baseline_results}
\end{table*}

In \cref{fig_EG_ell_measurement} we show the measurement of $E_G^{\ell}$ as a function of scale. These are calculated using the $E_G$ estimator of \cref{E_G_estimator}, using the measured angular cross-power spectra between the ACT DR6 lensing map and the SDSS BOSS DR12 CMASS and LOWZ galaxy samples together with the angular auto-power spectra of the galaxy samples and $\beta$ from an RSD analysis of the galaxy samples. We account for the cross-covariance between the cross- and auto-power spectra and calculate the PDF for $E_G$ without assuming Gaussianity in the result. The Figure shows the median of the PDF for each bin in angular scale and the uncertainty is based on the central 68.27\% confidence interval of the PDF. The constraints are shown for three different estimates of the uncertainty. The leftmost in orange are the results with the simulation-based covariance, in the middle in blue the constraints using the analytic covariance, and on the right in green the constraints using a jackknife-based covariance. In \cref{sec_covariance_estimation} we described in detail how the simulation-based covariance and analytic covariance are consistent within statistical expectations. Here they give consistent results for $E_G^{\ell}$ within the statistical accuracy of the simulation-based covariance. The jackknife-based covariance serves as a consistency check. It is a systematic overestimate of the uncertainty of 20-30\% and has large statistical uncertainty as it is based on only a small number of subsamples. Its advantage is that it is based on the data itself and could reveal additional sources of uncertainty not captured in the other methods. We here again find it to be consistent with the other methods within statistical expectations giving confidence to the accuracy of the approach. Finally \cref{fig_EG_ell_measurement} also shows the $\Lambda$CDM prediction for reference. A key feature of the GR-based predictions for $E_G$ is that on linear scales the statistic is expected to be scale-independent. Therefore, before comparing the overall amplitude with the $\Lambda$CDM-based prediction we can more generally test if the measurement is consistent with scale independence. We perform a PTE value test for consistency of the measurement with a scale-independent constant. The constant here is a free parameter not fixed to the $\Lambda$CDM prediction. We find $\textrm{PTE}_{\operatorname{scale-indep}}=0.74$ for ACT + CMASS and $\textrm{PTE}_{\operatorname{scale-indep}}=0.95$ for ACT + LOWZ and therefore no statistical evidence for a deviation from scale-independence within the accuracy of the measurement. For ACT + CMASS 9 out of the 11 bins are below the GR prediction. However, the different bins are significantly correlated so to evaluate if this is statistically significant we combine the measurements under consideration of the full covariance.  

\begin{figure}
\includegraphics[width=\columnwidth]{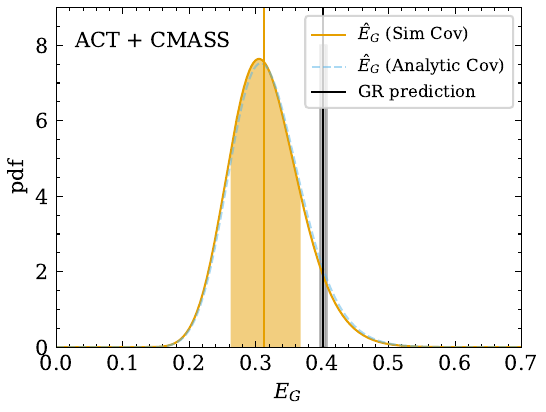}
\includegraphics[width=\columnwidth]{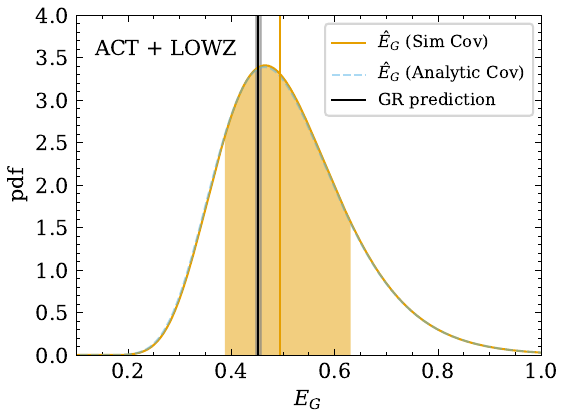}
\caption{We show the PDF for the combined constraints on $E_G$ over the cosmological analysis ranges. [Upper] The result for ACT and CMASS and [Lower] for ACT and LOWZ. In each case, the PDF for the combined constraint is shown using the simulation-based covariance [orange] and analytic covariance [blue, dashed]. For the former, the central 68.27\% confidence intervals are shaded in orange and the median is highlighted as a vertical line. The measurements can be compared to the GR prediction for the $E_G$ statistic at the effective redshift of the measurement based on $\Lambda$CDM constraints [vertical black line] with the $1\sigma$ uncertainty of the predicted value shaded. }
\label{fig_EG_measurement_pdf}
\end{figure}

Since the measurement is consistent with a constant we combine the constraints across the cosmological analysis range into an overall constraint on $E_G$. For this, we account for the full covariance between $C_\ell^{\kappa g}$ and $C_\ell^{g g}$ (see \cref{sec_cross_covariance}). For the jackknife-based uncertainty estimate, there are not enough samples to invert this full covariance, therefore these combined constraints are calculated using the simulation-based and the analytic covariances. \cref{fig_EG_measurement_pdf} shows the PDFs for the overall $E_G$ constraints. The $E_G$ statistic is measured for the effective redshift of the galaxy sample as given by \cref{eq_effective_redshift}, which results in $z_{\rm eff} = 0.555$ for CMB lensing combined with CMASS and in $z_{\rm eff} = 0.316$ for combinations with LOWZ. At these redshifts the expected values for the $E_G$ statistic for our fiducial cosmology assuming $\Lambda$CDM are given by $E_G^{\rm GR} (z_{\rm eff} = 0.555) = 0.401\pm 0.005$ and $E_G^{\rm GR} (z_{\rm eff} = 0.316) = 0.452\pm0.005$. These predictions are shown as black vertical lines in \cref{fig_EG_measurement_pdf}. The Gaussian PDF for the predictions is too highly peaked to show in the same plot so the $1\sigma$ uncertainty for the predicted values is shown as a gray band instead. These predictions can be compared with the measurements based on ACT + CMASS and ACT + LOWZ. In each case, the PDF using the simulation-based covariance and the analytic covariance are shown. The results using either covariance are consistent. For the simulation-based covariance, the median is highlighted with a vertical orange line, and the central 68.27\% confidence interval is shaded in orange. These result in $E_G^{\rm ACT+CMASS} = 0.31^{+0.06}_{-0.05}$ and $E_G^{\rm ACT+LOWZ} = 0.49^{+0.14}_{-0.11}$ respectively. We note that the PDFs, especially for the LOWZ case, are significantly non-Gaussian and result in asymmetric uncertainties. The asymmetry is such that the error towards higher values is larger than towards lower values. 

We test the measured $E_G$ values for consistency with the $\Lambda$CDM prediction resulting in $\textrm{PTE}_{\rm GR} = 0.13$ for ACT + CMASS and $\textrm{PTE}_{\rm GR} = 0.71$ for ACT + LOWZ. Both measurements are within statistical expectations, showing no statistical deviation from the prediction based on $\Lambda$CDM. For ACT + CMASS the 68.27\% confidence range does not include the prediction, i.e. the measurement differs by more than the equivalent of $1\sigma$ from the prediction. However, the deviation is not statistically significant with $\textrm{PTE}_{\rm GR} < 0.05$, meaning the 95\% confidence interval, approximately equivalent to the $2\sigma$ error of a Gaussian, includes the prediction. Therefore within the statistical constraining power of the measurements, the data shows no evidence for a deviation from the $\Lambda$CDM prediction. The $E_G$ results and PTE value tests are summarized in \cref{tab_baseline_results}. In \cref{sec_analysis_choices}, we evaluate if these conclusions are sensitive to analysis choices. For an overview of $E_G$ values in alternative gravity scenarios see \citet{Pullen2015}.

In addition to the statistical uncertainty we also estimate the systematic bias in the results. We combine in quadrature the conservative systematic bias estimates from galaxy bias evolution with redshift (\cref{sec_EG_gravity_statistic}), the level at which we validated the analysis pipeline (\cref{sec_input_recovery}) and the upper limit on the effective galaxy bias difference between the full galaxy maps and the overlap with ACT DR6 lensing. We find systematic error budgets of 3\% for ACT + CMASS and 4\% for ACT + LOWZ. These are significantly smaller than the statistical uncertainties of the measurement (approximately 18\% and 26\% respectively) and therefore negligible when evaluating the consistency of the measurement with GR predictions.

\subsection{Sensitivity to analysis choices} \label{sec_analysis_choices}

\begin{figure*}
\includegraphics[width=1.4\columnwidth]{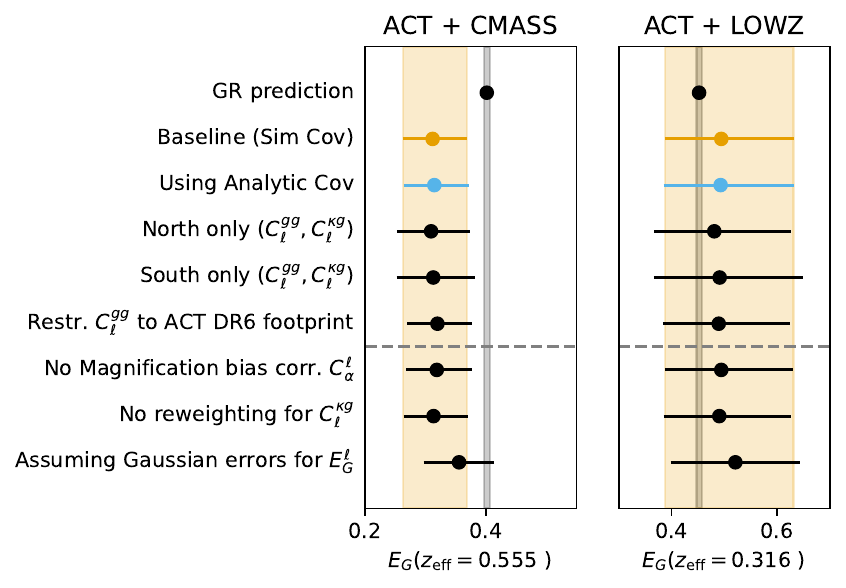}
\caption{Evaluating the sensitivity of the $E_G$ measurements to analysis choices for [Left] ACT + CMASS and [Right] ACT + LOWZ. The GR prediction [gray band] and baseline measurement using the simulation-based covariance [orange band] are used as references. These can be compared with a range of possible analysis choices: using an analytic covariance, calculating the angular power spectra only for the North or South patches, and restricting the $C_\ell^{gg}$ calculation to the overall with the ACT DR6 lensing map footprint. Furthermore, we also show the impact of removing well-justified analysis steps including the magnification bias correction, the reweighting for the angular cross-power spectrum, and assuming Gaussianity for $E_G$. For each case, the 68.27\% confidence ranges are shown. For a discussion see \cref{sec_analysis_choices}. }
\label{fig_analysis_choices_impact}
\end{figure*}

To evaluate the robustness of the presented $E_G$ results (\cref{sec_EGresults}) we investigate the sensitivity of the results to several analysis choices. These analysis choices were fixed before unblinding the data and the sensitivity to these analysis choices was only evaluated after the baseline results were finalized. See \cref{sec_blinding} for a detailed description of the blinding approach. 

In \cref{fig_analysis_choices_impact} the baseline results and GR predictions are compared to these alternative analysis choices for both the ACT + CMASS and ACT + LOWZ cases. For each case considered the median of the PDF and 68.27\% confidence intervals are compared to the GR prediction and baseline results. The tests include alternative analysis choices that could have been reasonably chosen for the analysis. Strong outliers for these could point to issues in the analysis approach or systematic effects that were not considered. We also consider what happens if some of the well-justified analysis steps we do include are excluded from the analysis. Here the results can be affected significantly and we investigate them for reference to demonstrate how much our conclusions depend upon the correctness of these analysis steps, or if their impact is small compared to the statistical uncertainty in the measurement. 

As shown in \cref{fig_EG_measurement_pdf} the PDF using the analytic covariance instead of the simulation-based covariance gives consistent results. This also results in the median and 68.27\% confidence intervals agreeing closely with the baseline simulation-based covariance. This agreement between different covariance estimates further shows the robustness of the analysis. Furthermore, since the two are consistent we decided to use the analytic covariance for combining the $E_G$ measurements using ACT and \textsl{Planck} for numerical accuracy.

In this analysis, we report combined constraints from the SDSS BOSS North and South patches after testing the angular cross-power spectra for each region for consistency (see \cref{sec_north_v_south}). For completeness, we also show the $E_G$ results for the angular power spectra calculated only for the North or South patches. Note that the $\beta$ values for the full maps are used, as the BOSS data vectors for the RSD analysis are only available for the full map \citep{Wenzl2024_EGestimator}. Except for $\beta$, the two measurements are independent and therefore can differ statistically. We find them consistent with expectations. 

For the baseline analysis, the full galaxy maps are used to calculate $C_\ell^{gg}$. In \cref{sec_galaxy_bias_diff} we evaluated the angular auto-power spectrum for consistency between the full maps and the overlap to estimate a systematic error budget. We here for completeness show the $E_G$ result for the alternative analysis choice of restricting the $C_\ell^{gg}$ to the footprint of the ACT DR6 maps. As the two $C_\ell^{gg}$ measurements are consistent the resulting $E_G$ values are also closely consistent. For additional discussion of the $C_\ell^{gg}$ measurements used as part of the $E_G$ estimate, we refer the interested reader to \citet{Wenzl2024_EGestimator}. 

Apart from these analysis choices, we can also evaluate the impact of parts of the analysis. We correct for the effect of magnification bias in our measurement (\cref{sec:magnification_bias}). Below the gray dashed line in \cref{fig_analysis_choices_impact} the results for $E_G$ without this correction are shown. For ACT + LOWZ the magnification bias effect is negligible. For ACT + CMASS the correction is 2\textendash 3\%, a small difference compared to the statistical uncertainty. The overall conclusions of the analysis are not affected by the magnification bias correction. For future surveys with larger statistical constraining power, the correction will become increasingly impactful.

We also test the effect of not using the additional reweighting for the cross-correlation. In this case, the effective redshift of the cross-correlation mismatches slightly with the effective redshift of the auto-correlation and $\beta$. This causes a scale-dependent percent level systematic for the $E_G$ estimate \citep{Wenzl2024_EGestimator}. This bias is small compared to the statistical uncertainty of current data. 

Finally, we show the results when assuming Gaussianity in the $E_G$ result. Since the PDF for our $E_G$ measurements is significantly non-Gaussian this can affect the resulting constraint. Based on the arguments laid out in \citet{Wenzl2024_EGestimator} the non-Gaussianity of the estimator is accounted for in our baseline results. The results when assuming Gaussianity are more directly comparable with previous analyses that did assume Gaussianty. We find that the measured values are affected but that the overall conclusion that the measurement is consistent with the $\Lambda$CDM-based prediction remains the same.  

\subsection{Combining ACT and Planck}

\begin{figure}
\includegraphics[width=\columnwidth]{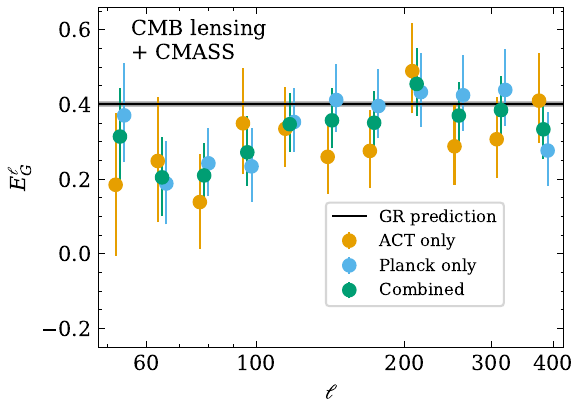}
\includegraphics[width=\columnwidth]{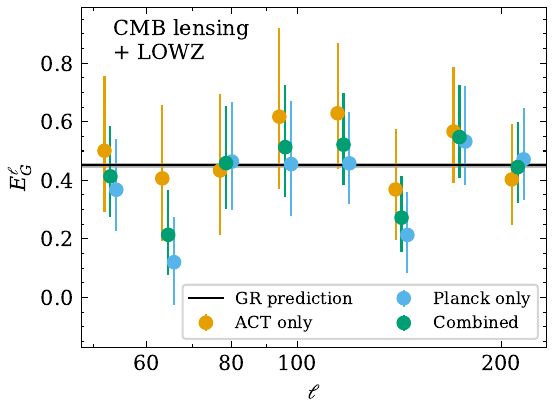}
\caption{Combined measurements of $E_G^{\ell}$ as a function of scale using ACT and \textsl{Planck} CMB lensing information. [Upper] The results with CMASS and [Lower] with LOWZ. In each case, we show the results using ACT alone [orange, left], \textsl{Planck} [blue, right], and the combined measurement [green, center]. The results shown are based on the analytic covariance estimate for numerical stability and the results are slightly shifted left and right for visibility. }
\label{fig_EG_ell_combined}
\end{figure}

\begin{figure}
\includegraphics[width=\columnwidth]{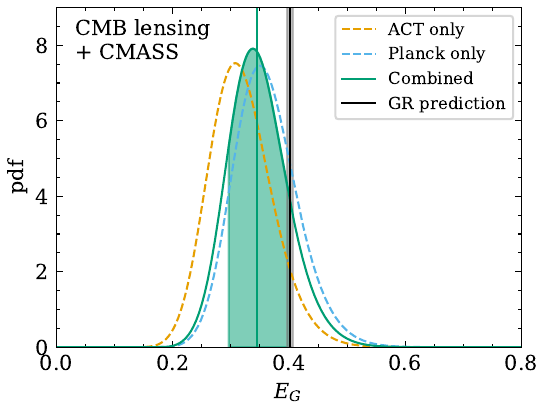}
\includegraphics[width=\columnwidth]{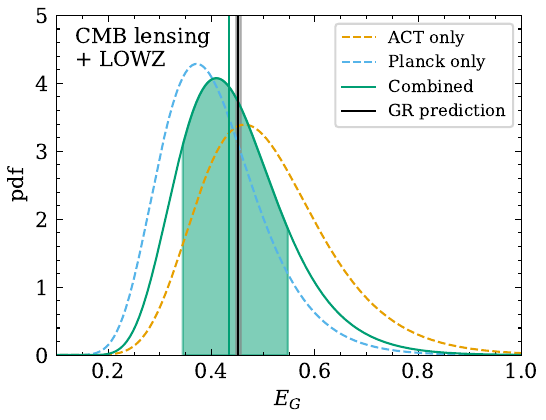}
\caption{PDFs for the measurement of $E_G$ using ACT and \textsl{Planck} CMB lensing information. [Upper] The result for CMASS and [Lower] for LOWZ. In each case, the PDF for ACT alone [orange, dashed line], \textsl{Planck} alone [blue, dashed], and the combined constraint [green line, median highlight with green vertical line] is shown. These measurements are based on the analytic covariance for numerical stability. They can be compared to the $\Lambda$CDM prediction [vertical black line] with $1\sigma$ uncertainty band shaded. }
\label{fig_EG_measurement_combined}
\end{figure}

We can combine the measurements based on ACT DR6 CMB lensing and \textsl{Planck} PR4 CMB lensing to obtain the strongest constraints on $E_G$ based on CMB lensing and galaxy clustering to date.

In \cref{sec_combined_cross_correlation} we discussed in detail how we calculated the minimum variance combination of the angular cross-power spectra at the bandpower level, $C_\ell^{\kappa g,\rm comb}$, and estimated the covariance for the combination. To estimate this combination numerically accurate we here switch to using the analytic covariance matrices that we have confirmed to give consistent results (see \cref{sec_covariance_estimation,sec_analysis_choices}).

In order to calculate $E_G$ we also need the cross-covariance between the $C_\ell^{\kappa g,\rm comb}$ measurement and the auto-power spectrum of the SDSS samples $C_\ell^{gg}$. The analytic calculation for this cross-covariance only depends on the masks of the $\kappa$ and galaxy maps, the cross-power spectrum, and the auto-power spectrum of the galaxies including noise. Crucially, the noise level of the CMB lensing map does not contribute to this cross-covariance as it is not correlated with the galaxy map and therefore the auto-power spectrum. Since the area covered by the ACT CMB lensing map is approximately a subset of the area of the \textsl{Planck} CMB lensing map we use the analytic cross-covariance between \textsl{Planck} CMB lensing map and the SDSS BOSS samples presented in \citet{Wenzl2024_EGestimator} as an estimate for the cross-covariance.

In \cref{fig_EG_ell_combined} we show the results for $E_G^{\ell}$ from CMB lensing data with CMASS and LOWZ from ACT and \textsl{Planck} alone and from ACT + \textsl{Planck} combined. The two individual measurements are strongly correlated, showing similar trends with scale. Specifically, we find the same trend of the best fit $E_G^{\ell}$ values being lower than the theory prediction on large scales for CMB lensing + CMASS as noted in previous harmonic space analyses \citep{Pullen2016,Wenzl2024_EGestimator}. We test our most constraining combined measurement of ACT + \textsl{Planck} + CMASS for consistency with scale independence finding $\textrm{PTE}_{\operatorname{scale-indep}} = 0.53$. Therefore just like for the individual measurements, the visual trend is not statistically significant for our measurement. Similarly, we find the ACT + \textsl{Planck} + CMASS measurement to also be consistent with scale-independence with $\textrm{PTE}_{\operatorname{scale-indep}} = 0.46$.

Since the combined constraint also does not show evidence for scale dependence we combine the bins in the cosmological range for overall constraints on $E_G$. We find $E_G^{\rm ACT+Planck+CMASS} = 0.34^{+0.05}_{-0.05}$ and $E_G^{\rm ACT+Planck+LOWZ} = 0.43^{+0.11}_{-0.09}$. The combined results sample the same effective redshift as the ACT-only results and therefore have the same expected values. Testing them for consistency with the measurement we find $\textrm{PTE}_{\rm GR} = 0.31$ for ACT + \textsl{Planck} + CMASS and $\textrm{PTE}_{\rm GR} = 0.86$ for ACT + \textsl{Planck} + LOWZ. Therefore the combined measurements show no evidence for a statistical deviation from $\Lambda$CDM predictions. The combined results are summarized together with the ACT-only results in \cref{tab_baseline_results}.

\subsection{Results in context} \label{sec_results_in_context}

\begin{figure}
\includegraphics[width=\columnwidth]{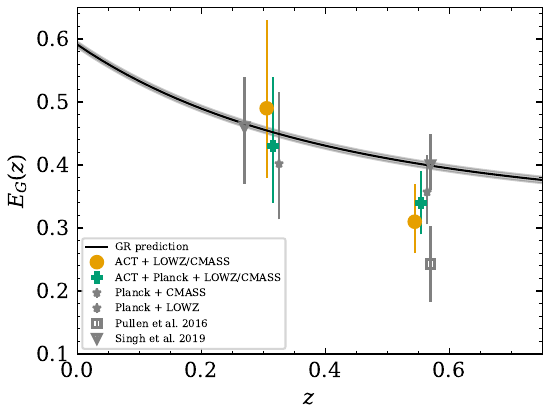}
\includegraphics[width=\columnwidth]{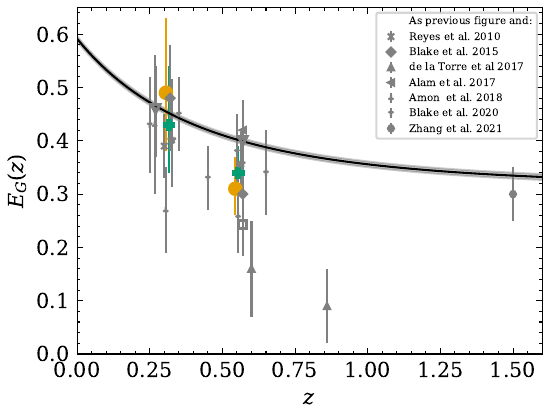}
\caption{Comparison of the $E_G$ measurements presented in this work with a selection of literature results using CMB lensing and galaxy information (top panel) and all $E_G$ measurements to date (bottom panel). Shown are the results using ACT + LOWZ or CMASS [orange circle, shifted left slightly], ACT + \textsl{Planck} + CMASS or LOWZ [green plus]. These can be compared with the \textsl{Planck} only results [gray star, shifted right slightly] which were consistently analyzed in \citet{Wenzl2024_EGestimator}. Furthermore, they can be compared with the results presented in \citet{Singh2019} [gray square] and \citet{Pullen2016}. The prediction for $E_G$ based on $\Lambda$CDM is shown as a black line with a shaded band for the $1\sigma$ uncertainty. The bottom panel shows additional measurements at higher redshift as well as from using galaxy lensing. See \cref{sec_results_in_context} for a discussion. }
\label{fig_literature_comparison}
\end{figure}

We can compare the $E_G$ measurements presented in this work to measurements available in the literature. In \cref{fig_literature_comparison} we show the published measurements using CMB lensing and galaxy clustering in the redshift range $0<z<1$ which are most directly comparable to the results presented in this work. 

All of the previous measurements shown in \cref{fig_literature_comparison} are based on SDSS BOSS data, specifically the CMASS and LOWZ galaxy samples also used in this work, and CMB lensing information from \textsl{Planck}. As discussed in \citet{Wenzl2024_EGestimator} the difference between the \textsl{Planck} + CMASS and \textsl{Planck} + LOWZ measurements to the \citet{Singh2019} results is small and can reasonably be attributed to a different in the effective redshift for LOWZ due to a different redshift cut on the galaxy sample and the simplifying assumption of Gaussianity in $E_G$ made for the previous analysis. Neither measurement reproduced the early $E_G$ detection from \citet{Pullen2015} which was based on earlier versions of the SDSS BOSS and \textsl{Planck} CMB lensing data. For detailed discussions see \citet{Wenzl2024_EGestimator} and \citet{Singh2019}.

The ACT DR6 lensing map used in this analysis gives constraints on $E_G$ independent of \textsl{Planck}. In \cref{fig_literature_comparison} the \textsl{Planck} based constraints can be compared to the ACT + CMASS [Orange circle] and ACT + LOWZ [Orange pentagon] constraints. Both measurements match the consistently analyzed \textsl{Planck} results from \citet{Wenzl2024_EGestimator} within the $68.27\%$ confidence range. Furthermore, the cosmological interpretation of both measurements is also consistent with neither showing statistical evidence for tension with the $\Lambda$CDM prediction. The $E_G$ analysis is overall limited by the uncertainty of the $C_\ell^{\kappa g}$ measurement, making this a strong consistency check between the \textsl{Planck} and ACT CMB lensing-based results. We can further compare the results to previous analyses using \textsl{Planck} CMB lensing data. The result from \citet{Singh2019} using \textsl{Planck} + LOWZ is consistent within the $68.27\%$ confidence range and the \textsl{Planck} + CMASS results differ by slightly more than the $68.27\%$ confidence range. We note that if we assumed Gaussianity for the result as done in \citet{Singh2019} then both measurements would match within the $68.27\%$ confidence range (see the last line of  \cref{fig_analysis_choices_impact}). The ACT + CMASS measurement result is approximately in between the low \textsl{Planck} + CMASS measurement reported in \citet{Pullen2016} and the other published measurements \citep{Singh2019,Wenzl2024_EGestimator}. It does not disagree significantly with any of the three measurements. Notably, however, the result is not in tension with the $\Lambda$CDM predictions similar to \citep{Singh2019,Wenzl2024_EGestimator} whereas the \citet{Pullen2016} result does show a larger $2.6\sigma$ difference with the prediction. 

Since the measurements based on ACT presented here and the consistent analysis results of \citet{Wenzl2024_EGestimator} using \textsl{Planck} are statistically consistent we derive combined constraints on $E_G$. We carefully estimate the uncertainties in this analysis, correctly accounting for the non-Gaussianity of the measurements. This resulted in combined errors whose uncertainties are not significantly smaller than those reported in \citep{Pullen2016,Singh2019}, (which did not account for this non-Gaussianity). However, the combined constraints contain the most information of all measurements to date and have smaller uncertainty than the consistently analyzed individual constraints using either survey. Therefore we refer to the reported constraints as the strongest constraints on $E_G$ using CMB lensing and galaxy clustering to date. We find, as shown in \cref{fig_literature_comparison}, that both the LOWZ- and CMASS-based constraints show statistical consistency with the $\Lambda$CDM predictions and sit in between the ACT only and \textsl{Planck} only constraints.

Another measurement of $E_G$ at higher redshift has been performed in \citet{Zhang2021} using \textsl{Planck} CMB lensing and SDSS quasars. They also reported consistency with the same $\Lambda$CDM fiducial cosmology, probing $E_G$ at an effective redshift of $z=1.5$. Furthermore, $E_G$ can also be estimated using weak lensing of galaxies instead of CMB lensing. An overview of such measurements can be found in Fig. 11 of \citep{Zhang2021}. The majority of these measurements also find consistency with the $\Lambda$CDM prediction \citep{Reyes2010,Blake2016,Alam2017,Singh2019,Blake2020}, with only some reporting low outliers \citep{delaTorre2017,Amon2018}.

In this analysis for the combined ACT+\textsl{Planck} cross-correlation measurement, we assume the noise in the ACT and \textsl{Planck} CMB lensing maps is uncorrelated with each other. As discussed in \citet{Qu2023}, a noise contribution arising from the reconstruction is correlated between the CMB lensing maps. As shown in Figure 6 of \citet{Qu2023} this reconstruction noise is a small fraction of the noise+cosmic variance for $\ell \leq 200$. On smaller scales, $200<\ell <433$, as used for CMASS, this correlated reconstruction noise can become significant. By considering a worst-case scenario that the correlated noise is equal to the full noise we obtain a bound on the potential impact of reconstruction noise for the $E_G$ statistic. We find that our final combined $E_G$ constraint would be shifted by 0.0025 (0.7\% of the result of $E_G = 0.34$) and the upper and lower uncertainty would be shifted by 0.0007 and 0.0006 (both 1.2\% of the uncertainty of $\pm0.05$) respectively. The effect of the reconstruction noise is well within our systematic error budget and therefore negligible in this analysis. This additional contribution could be captured in future work using simulations similar to what was done in \citet{Qu2023}.

Overall the differences in results for $E_G$ from various analyses with very similar datasets probing the same statistic show the need for robust and careful analyses to derive reliable conclusions. This will become more relevant for upcoming datasets with increased statistical constraining power so that systematic effects become increasingly relevant.

\section{Conclusion} 
\label{sec_conclusion}

In this work, we presented a measurement of the cross-correlation between the ACT DR6 lensing map and the SDSS BOSS galaxy samples CMASS and LOWZ and validated the analysis pipeline. We then used these measurements to constrain the $E_G$ statistic of gravity, independent from \textsl{Planck}. Furthermore, we derived combined constraints leveraging both CMB lensing maps to report the strongest constraints on $E_G$ using CMB lensing and galaxy clustering to date. We found that both the measurements with the independent ACT data and the combined constraints are statistically consistent with $\Lambda$CDM predictions.

The cross-correlation measurement was developed following the ACT DR6 lensing blinding guidelines. Through a set of null and consistency checks, we validated the accuracy of the analysis approach before unblinding the results. We showed that the analysis pipeline can accurately recover the input within statistical expectations for the set of 400 simulated map realizations. We also showed consistency between three different covariance estimates using this set of simulations, an analytic calculation, and a jackknife approach. We showed that the bias-hardened estimator used for the baseline ACT DR6 CMB lensing map sufficiently mitigates the effect of extragalactic foregrounds for the example of the \websky simulation. We performed difference tests between the $90~\textrm{GHz}$ and $150~\textrm{GHz}$ measurements of the ACT CMB observations to further test for contamination from foregrounds finding them to be consistent. Finally, we also showed that an alternative approach using CIB-deprojection to mitigate extragalactic foreground gives consistent results to our baseline. Together with additional tests that purposefully misalign the maps to correlate and a test to show the consistency of the North and South patches individually, we concluded that the analysis approach is robust. The distribution of all PTE value tests reported is within statistical expectations giving further confidence to the statistics evaluated. 

Applying this carefully developed analysis pipeline to the data we reported new cross-correlation measurements between ACT DR6 CMB lensing and the SDSS BOSS CMASS and LOWZ galaxy samples (\cref{fig:cross_corr_results}). For the cosmological analysis, we only used large scales where one can assume linear theory. In addition, we also validated and reported the cross-correlation measurement to smaller scales where additional data is available that can be leveraged in future analyses. We make these measurements available upon publication of the work which can be used to gain additional constraining power from the data when modeling smaller scales.

Based on the new constraints on the cross-correlation we derived estimates of the $E_G$ statistic of gravity. To estimate the statistic we additionally used previous analyses of the SDSS BOSS samples, specifically we used the galaxy auto-correlation and $\beta$ results reported for CMASS and LOWZ in \citet{Wenzl2024_EGestimator}. The resulting values for $E_G$ were summarized in \cref{tab_baseline_results}. We tested our $E_G$ measurement both for consistency with scale independence as predicted by GR generally and tested the overall constraint on $E_G$ against the predicted values assuming $\Lambda$CDM based on fits to primary CMB anisotropies and BAO. We found both the ACT only and combined constraints statistically consistent with the predictions for our fiducial cosmology. This indicates that within our measurement uncertainty, the observed growth of structure is consistent with predictions using $\Lambda$CDM fitted to the background expansion and early universe. Our combined measurements achieve fractional errors of approximately 15\% with ACT and \textsl{Planck} CMB lensing together with CMASS and 23\% with LOWZ. Our results are in line with a wide range of investigations of $E_G$ in the literature most of which found similar consistency with GR predictions \citep[for a recent overview see ][]{Zhang2021}. The developed techniques to measure $E_G$ are designed to be unbiased to a much higher level with systematic errors of only 3\% for ACT + CMASS and 4\% for ACT + LOWZ. This systematic error is dominated by the conservative accounting for a potential effective bias difference between the full galaxy maps and the overlap with ACT which could be avoided in future work by recalculating the RSD analysis for the specific overlaps with CMB lensing maps. The main astrophysical systematic of galaxy bias evolution with redshift can also be mitigated further for analyses with larger datasets by calculating $E_G$ in narrower bins in redshift. Therefore future data with larger statistical constraining power offers the opportunity to improve upon these constraints and probe gravity at the percent level. This will allow discriminating even smaller deviations from the predictions for the growth of the structure in alternative gravitational scenarios and General Relativity \citep{Pullen2015}.

The analysis techniques developed in this work are directly applicable to future investigations of the $E_G$ statistic. The systematic corrections developed will become increasingly relevant as the statistical uncertainties decrease with further data. Of considerable interest will be combining the ACT DR6 CMB lensing map with upcoming spectroscopic galaxy samples from DESI. Beyond this, a range of upcoming surveys will produce new competitive CMB lensing maps including SO and CMB-S4 as well as upcoming galaxy samples from SPHEREx, \textsl{Euclid}, LSST, and Roman. The latter surveys will not have spectroscopic redshift information available, making estimating $\beta$ accurately challenging but through their significantly larger galaxy samples would allow tight constraints on the angular power spectra \citep{Pullen2015}.

\begin{acknowledgments}

We thank Martin White for helpful discussions on best practices for HODs from SDSS BOSS publications. We thank Sigurd K. Næss and Bruce Partridge for helpful comments on a draft of this paper. 

The work of LW and RB is supported by NSF grant AST-2206088 and NASA ROSES grant 12-EUCLID12-0004. 
SC acknowledges the support of the National Science Foundation at the Institute for Advanced Study through NSF/PHY 2207583. 
GSF acknowledges support through the Isaac Newton Studentship and the Helen Stone Scholarship at the University of Cambridge. GSF, FJQ, BS, and NM acknowledge support from the European Research Council (ERC) under the European Union’s Horizon 2020 research and innovation programme (Grant agreement No. 851274). 
GAM is part of Fermi Research Alliance, LLC under Contract No. DE-AC02-07CH11359 with the U.S. Department of Energy, Office of Science, Office of High Energy Physics. 
MM acknowledges support from NSF grants AST-2307727 and  AST-2153201 and NASA grant 21-ATP21-0145. 
AvE acknowledges support from NASA grants 80NSSC23K0747 and 80NSSC23K0464. 
CS acknowledges support from the Agencia Nacional de Investigaci\'on y Desarrollo (ANID) through Basal project FB210003. 
IH acknowledges support from the European Research Council (ERC) under the European Union's Horizon 2020 research and innovation programme (Grant agreement No. 849169). 
OD acknowledges support from a SNSF Eccellenza Professorial Fellowship (No. 186879). 
EC acknowledges support from the European Research Council (ERC) under the European Union’s Horizon 2020 research and innovation programme (Grant agreement No. 849169).

Support for ACT was through the U.S.~National Science Foundation through awards AST-0408698, AST-0965625, and AST-1440226 for the ACT project, as well as awards PHY-0355328, PHY-0855887 and PHY-1214379. Funding was also provided by Princeton University, the University of Pennsylvania, and a Canada Foundation for Innovation (CFI) award to UBC. ACT operated in the Parque Astron\'omico Atacama in northern Chile under the auspices of the Agencia Nacional de Investigaci\'on y Desarrollo (ANID). The development of multichroic detectors and lenses was supported by NASA grants NNX13AE56G and NNX14AB58G. Detector research at NIST was supported by the NIST Innovations in Measurement Science program. Computing for ACT was performed using the Princeton Research Computing resources at Princeton University, the National Energy Research Scientific Computing Center (NERSC), and the Niagara supercomputer at the SciNet HPC Consortium. SciNet is funded by the CFI under the auspices of Compute Canada, the Government of Ontario, the Ontario Research Fund–Research Excellence, and the University of Toronto. We thank the Republic of Chile for hosting ACT in the northern Atacama, and the local indigenous Licanantay communities whom we follow in observing and learning from the night sky.

This research used resources of the National Energy Research Scientific Computing Center (NERSC), a U.S. Department of Energy Office of Science User Facility located at Lawrence Berkeley National Laboratory, operated under Contract No. DE-AC02-05CH11231 using NERSC award HEP-ERCAPmp107

The sky simulations used in this paper were developed by the WebSky Extragalactic CMB Mocks team, with the continuous support of the Canadian Institute for Theoretical Astrophysics (CITA), the Canadian Institute for Advanced Research (CIFAR), and the Natural Sciences and Engineering Council of Canada (NSERC), and were generated on the Niagara supercomputer at the SciNet HPC Consortium (cite https://arxiv.org/abs/1907.13600). SciNet is funded by: the Canada Foundation for Innovation under the auspices of Compute Canada; the Government of Ontario; Ontario Research Fund - Research Excellence; and the University of Toronto.

SDSS-III is managed by the Astrophysical Research Consortium for the Participating Institutions of the SDSS-III Collaboration including the University of Arizona, the Brazilian Participation Group, Brookhaven National Laboratory, Carnegie Mellon University, University of Florida, the French Participation Group, the German Participation Group, Harvard University, the Instituto de Astrofisica de Canarias, the Michigan State/Notre Dame/JINA Participation Group, Johns Hopkins University, Lawrence Berkeley National Laboratory, Max Planck Institute for Astrophysics, Max Planck Institute for Extraterrestrial Physics, New Mexico State University, New York University, Ohio State University, Pennsylvania State University, University of Portsmouth, Princeton University, the Spanish Participation Group, University of Tokyo, University of Utah, Vanderbilt University, University of Virginia, University of Washington, and Yale University.

\end{acknowledgments}

\appendix

\section{Combining the cross-correlation measurements} \label{combining_cross_corr_measurements}

Our procedure to calculate the combined angular cross-power spectrum of the measurements based on ACT $\times$ SDSS and \textit{Planck} $\times$ SDSS was as follows. We combine the measurements at the bandpower level using inverse variance weighting. Importantly the ACT and \textit{Planck} CMB lensing maps overlap and therefore, the cross-correlation measurements have a non-negligible correlation. In the overlap region between the \textit{Planck} and ACT CMB lensing maps the cosmic variance is correlated and both cross-correlation measurements use the same galaxy samples. We need to account for this correlation when estimating the covariance of the combined measurement. 

The inverse variance weighted combination of two measurements $A, B$ with covariances $\textrm{Cov}_A$ and $\textrm{Cov}_B$ is given by
\begin{align}
    C &= H (\textrm{Cov}_A^{-1} A + \textrm{Cov}_B^{-1}  B) \label{inverse_variance_weighted_combination}, \\ 
    H &\equiv \left(\textrm{Cov}_A^{-1} + \textrm{Cov}_B^{-1}\right)^{-1}.
\end{align}
If the two measurements are independent then the inverse covariance for the combination \cref{inverse_variance_weighted_combination} is given by
\begin{align}
    \textrm{Cov}^{-1} = (\textrm{Cov}_A^{-1} + \textrm{Cov}_B^{-1}). \label{combined_covariance_independent}
\end{align}
This approach could be applied to our data if we cut out part of the \textit{Planck} map to force the correlations to be independent. This can be achieved by cutting out the ACT region from the \textit{Planck} map and combining ACT $\times$ SDSS with the correlation of the area-restricted \textit{Planck} map with SDSS. To calculate the combined inverse covariance we only need unbiased estimates of the inverse covariance for each measurement allowing us to apply this to noisy estimates of the covariance, like a simulation-based covariance estimate. 

If the two measurements $A, B$ are not independent then we need to account for the cross-covariance $\textrm{CrossCov}_{AB}$ between the measurements. Error propagation for the combination in \cref{inverse_variance_weighted_combination} then gives
\begin{align}
    \textrm{Cov} =& H (I + \textrm{Cov}_A^{-1} \textrm{CrossCov}_{AB} H \textrm{Cov}_B^{-1}  \nonumber \\  &+ \textrm{Cov}_B^{-1} \textrm{CrossCov}_{BA} H \textrm{Cov}_A^{-1} ) , \label{combined_covariance_correlated}
\end{align}
where $I$ is the identity matrix. For small cross-covariances, this converges back to \cref{combined_covariance_independent}. And for $\textrm{Cov}_A = \textrm{Cov}_B$ and $100\% $ correlation between the datasets the combination has no additional information and therefore $\textrm{Cov} = \textrm{Cov}_A = \textrm{Cov}_B$. 

The calculation in \cref{combined_covariance_correlated} is difficult to perform in practice for noisy estimates of the covariance as one needs to invert multiple times. We therefore only use the analytic covariance here to be able to estimate this reliably. 

\begin{figure}
\includegraphics[width=\columnwidth]{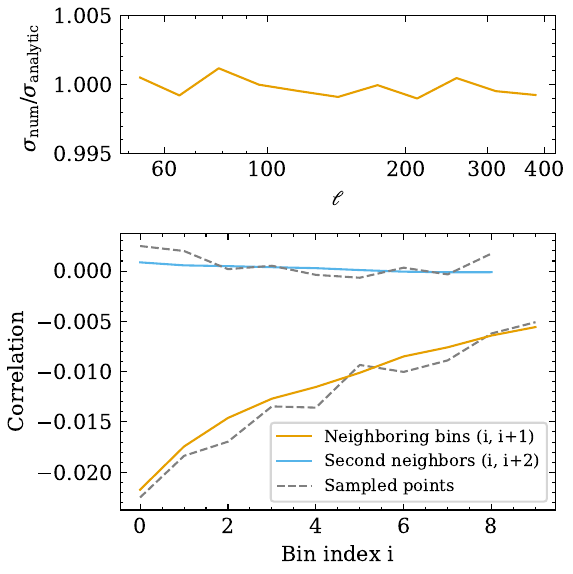}
\caption{Verification of our analytic combined covariance by comparing with the numerical combined covariance of 10 Million samples from the two covariances. The top plot shows the ratio of the marginalized uncertainty of the numerical estimate with the analytic estimate, they agree within $0.2\%$ percent. Additionally, we calculate the correlation matrix for the covariances. The bottom plot shows the analytic estimate of the correlation for neighboring bins [orange solid line] and second neighbors [blue line] compared with the results from the sampled covariance [gray dashed line].}
\label{fig:combined_covariance_convergence}
\end{figure}

We test the accuracy of \cref{combined_covariance_correlated} by comparing it to a large sampling of the covariances for the case of ACT $\times$ CMASS and \textit{Planck} $\times$ CMASS. To do this we take samples from a normal distribution ($\mathcal{N}$) with means for A and B given by our fiducial cosmology $C_\ell^{\kappa g}$ and using the measurement covariances given by
\begin{align}
    \mathcal{N} & \left( \mu =\begin{pmatrix}
C_\ell^{\kappa g} \\
C_\ell^{\kappa g} 
\end{pmatrix}, \right. \nonumber\\ & \left. \textrm{Cov} = \begin{pmatrix}
\textrm{Cov}_A & \textrm{CrossCov}_{AB} \\
\textrm{CrossCov}_{BA} & \textrm{Cov}_B
\end{pmatrix} \right).
\end{align}
We then combine each sample as given by \cref{inverse_variance_weighted_combination} and estimate the covariance numerically from the variance of the points, we use 10 Million samples. In \cref{fig:combined_covariance_convergence} we show a comparison of the diagonal of the combined covariance and a comparison of the correlation between the neighboring bins and with the second neighbors. For further apart bins the correlations become negligible. Our analytic estimate of the combined covariance (\cref{combined_covariance_correlated}) matches well with the numerical estimate. The estimate of the marginalized uncertainty agrees within $0.2\%$ and the off-diagonals are captured well within the percent level.

\bibliography{main}

\end{document}